\documentclass[acmsmall]{acmart}

\usepackage[flushleft]{threeparttable}
\usepackage{paralist}
\usepackage{adjustbox}
\usepackage{quoting,xparse,url,hyperref}

\hypersetup{breaklinks=true}

\NewDocumentCommand{\bywhom}{m}{
  {\nobreak\hfill\penalty50\hskip1em\null\nobreak
   \hfill\mbox{\normalfont(#1)}%
   \parfillskip=0pt \finalhyphendemerits=0 \par}%
}

\NewDocumentEnvironment{pquotation}{m}
  {\vskip1em\begin{quoting}[
     indentfirst=true,
     leftmargin=\parindent,
     rightmargin=\parindent]\itshape}
  {\bywhom{#1}\vskip1em\end{quoting}}

\AtBeginDocument{%
  \providecommand\BibTeX{{%
    \normalfont B\kern-0.5em{\scshape i\kern-0.25em b}\kern-0.8em\TeX}}}

\copyrightyear{2021}
\acmYear{2021}

\acmJournal{JACM}
\acmVolume{0}
\acmNumber{0}
\acmArticle{000}
\acmMonth{0}

\begin{document}

\title{Past, Present and Future of Computational Storage: A Survey}

\author{Corne Lukken}
\email{info@dantalion.nl}
\author{Animesh Trivedi}
\email{a.trivedi@vu.nl}
\affiliation{%
  \institution{Vrije Universiteit (VU)}
  \streetaddress{De Boelelaan 1105}
  \city{Amsterdam}
  \country{Netherlands}
  \postcode{1081 HV}
}

\renewcommand{\shortauthors}{Lukken and Trivedi}

\begin{abstract}
We live in a data-centric world where we are heading to generate close to 200 Zettabytes of data by the 
year 2025~\cite{2025-200-zb}. Our data processing requirements have also increased as we push to build data 
processing frameworks that can process large volumes of data in a short duration, a few milli- and even 
micro-seconds. In the prevalent computer systems designs, data is stored passively in storage devices which 
is brought in for processing and then the results are written out. As the volume of data explodes this 
constant data movement has led to a \textit{data movement wall} which hinders further process and optimizations 
in data processing systems designs. One promising alternative to this architecture is to push computation 
to the data (instead of the other way around), and design a computational-storage device or CSD. The idea 
of CSD is not new and can trace its root to the pioneering work done in the 1970s and 1990s. More recently, 
with the emergence of non-volatile memory (NVM) storage in the mainstream computing (e.g., NAND flash and Optane), 
the idea  has again gained a lot of traction with multiple academic and commercial prototypes being available now. 
In this brief survey we present a systematic analysis of work done in the area of computation storage and 
present future directions.   
\end{abstract}


\maketitle

\section{Introduction} 

Modern storage technologies such as flash or Intel optane \cite{intel-optane}
allow for increasingly higher bandwidth. Due to the Von Neumann architecture
this results in increasing amounts of data being moved between the storage
device, Central Processing Unit (CPU) and Dynamic Random Access Memory (DRAM).
However, not all these technologies are scaling at the same rate. Primarily CPU
and DRAM are projected to scale slower in the coming years. Other technological
improvements such as those of link and interconnect bandwidth might suddenly
stagnate\footnotemark[1]. The consequence is that excessive data movement is
rapidly becoming a bottleneck for data intensive
workloads \cite{2014-micro-ndp, cpu-improvements, 2018-neumann-bottleneck,
2016-western-digital}. Even if the technological capabilities keep scaling to
keep up with flash storage technology, unnecessarily moving data remains
wasteful.

\footnotetext[1]{The PCI-SIG group projected in 2019 that PCIe gen6 would be
available in 2021 \cite{pcie-predict-2019}. Currently a draft exists but no
matching electromechanical specification making it unclear if this theoretical
protocol will be feasible in practice. The combination of a closed drafting
process and the specifications being hidden behind a paywall make further
predictions impractical.}

A promising solution to this problem is pushing compute to the storage layer.
Today this is researched under a large variety of terms such as
"Near-data processing" or "Computational Storage". However, the concept of
Computational Storage (CS) is not new and has previously been explored for
mainframes \cite{database-computer}, distributed storage, databases, buses,
links, interconnects and harddrives (HDD) \cite{active-disks-tech,
active-disk-pillar}. The problem in identifying CS is that until recently an
exact definition was not available \cite{snia-model}. Throughout this work the
definitions set out by the SNIA Computational Storage Architecture and
Programming Model will be used \cite{snia-model}. This association is driving
research and standardization in storage and networking with a number of 
standards being drafted on Computational Storage. In addition SNIA hosts
several conferences.

In light of these recent definitions it is apparent that CS is ubiquitous, as
will be demonstrated in a later section. However, the adoption of CS varies
greatly across different types of storage technologies. Particularly in flash
based Computational Storage Devices (CSx) such as Solid State Drives (SSD).
Here we have seen nearly a decade of research with regards to CSx with no
widespread market adoption. In fact, it is only recently that the first
Programmable Flash Storage (PFS) Devicess have become commercially 
available \cite{product-samsung, product-newport, product-eideticom,
product-scaleflux}. The underlying flash architecture is substantially different
to that of convential harddrives while this providing the same block level
access interface\cite{death-interface}. This difference is known as the semantic
gap and is a potential reason for slow research progress in CS.

All currently available PFS devices either commercial or for research purposes
ignore fundamental requirements such as security, multi-user tenancy, usability
and data consistency \cite{barbalacecomputational}. Why is it that even after a
decade even the latest works do not propose solutions that meet these
requirements?





\section{Related Work}

We intentionally limit this section to surveys and notice that surveys on CS
are very limited. However, there is one key work from Antonio Barbalace \&
Jaeyoung Do \cite{barbalacecomputational}. Several other fields would likely
benefit from CS applications such as those of big data storage and
Software-Defined Storage (SDS) systems. SDS uses software abstractions to
provide provisioning and data management indenpent of the underyling hardware.
In these fields we see notable surveys that describe a similar problem of data
increase and performance requirements \cite{Siddiqa2017, 10.1145/3385896}.

Even the concept of programmability is directly addressed as an issue in SDS
systems. We believe the field of SDS and CS have many similarities but that SDS
does not typically employ computations close to the storage elements themselves.
Given the computations that are required for provisioning, policies and
management these could potentially be partially offloaded to the underlying
drives in the case of CS. This is a potential future opportunity in SDS to
utilize CS to improve energy efficiency and prevent unnecessary data movement.
Especially as many of SDS functions such as erase coding and data compression
are excellent CS applications.

Similarly in big data systems we see large scale databases and distributed
filesystems. In these applications we typically require fast retrieval and
realtime processing. Yet the architectures employed typically involves moving
the data over interconnects and even networks. Often the data retrieved is only
used once or temporarily and most applications such as those of MapReduce are
reductive in nature. Resulting in that these big data applications would also
likely benefit from utilizing CS.

While CS is able to help advance certain fields of research there is also work
being done that helps advance CS. Notable examples are the design of
host-managed interfaces for SSDs such as on Open-Channel SSD (OCSSD) and more
recently Zoned Namespaces (ZNS) \cite{10.1145/3448016.3457540}\footnotemark[2].
However, there is no general agreement that ZNS and CS are complementary
technologies and some even argue they are
contradictory \cite{10.1145/3458336.3465300}. In this study we will elaborate
extensively on why these two technologies work excellently together.

\footnotetext[2]{This reference refers to Computational Storage as if it solely
relates to black-box devices without any programmability, so called fixed
function devices. In this work we use the broader definition as set out by
the SNIA model \cite{snia-model}. additionally, the work its Computational
Devices section explains concepts of SNIA model 0.5 which has changed
significantly since then (0.8 at time of writing).}

Beyond the storage interface we see research in the programming interface as
well \cite{bpf-uapi}. In recent works a case is being made to use BPF for
heterogeneity. BPF is typically used within the context of the Linux kernel but
in essence it is an open Instruction Set Architecture (ISA). BPF bytecode can be
executed in a VM, presenting a stable Application Binary Interface (ABI) that is
completely decoupled from the host ISA and vendor implementation. Allowing for
reusable applications that can be written in any programming langauge as long as
they are compiled into BPF bytecode. This programming flexibility for users
combined with implementation flexibility for vendors is ideal for CS.

\section{Study Design}

In this section we describe our research goal as well as the questions that
support this. Afterwards we show a selection of initial papers and describe
a procedure to find related literature. Furthermore, we define the inclusion and
exclusion criteria for this literature as well as any exceptions. In addition,
we show an overview of the selected literature, the keywords used and a general
timeline from when these works where published. To finally, offering a dedicated
section on reproduceability.

\subsection{Research Goal}

Our study aims at creating a brief historical overview of CS given the new
definitions set out by SNIA \cite{snia-model}. From this, a case is made to
demonstrate the ubiquity of CS and all its current different applications.
Afterwards we draw our attention to PFS and thoroughly investigate all different
research prototypes over the past decade. Furthermore, we identify the
limitations of these prototypes and describe in detail why there is still no
general availability of PFS device. The main focus of this study is on the
subject of PFS. It is only on this subject that we provide future predictions and
suggestions.

From these goals the following survey questions are drawn:

\begin{enumerate}
    \item What is Computational Storage (CS)?
    \item What type of applications are there for CS?
    \item How has Programmable Flash Storage (PFS) advanced over the past decade?
    \item What are the challenges involving PFS?
    \item How will PFS evolve in the near future?
\end{enumerate}

\subsection{Seed Papers}

A large selection of research papers is needed to be able to completely answer
all the questions set out in our study design. However, an exhaustive approach
is simply not possible due to time constraints. Similarly the large variety of
keywords used throughout the history of CS makes a simple term based exploratory
study approach impractical.

Our approach relies on identifying a few initial seed papers to create a set of
keywords used for exploration. These keywords are not only extracted from the
seed papers themselves but also from any citations. Subsequently the keywords
are used to search across ACM, IEEE, SNIA, Usenix and Semantic Scholar.
The initial selection is done based on the definition for CS as set out
by SNIA \cite{snia-model}, however, it should be noted that this definition is
very broad. Filtering of this selection is done using several inclusion and
exclusion criteria that will be detailed later. In total 50 papers will be
selected of which 25 dedicated to PFS. This manual selection
procedure allows to identify the key historical contributions as well as the
state of the art while taking into account any time constraints.

Our selection of initial seed papers is shown sorted by publish date in
ascending order in table \ref{table:seedpapers}.

\begin{table}[h!]
	\caption{Seed Papers}
	\label{table:seedpapers}
	\centering
	\begin{adjustbox}{width=1\textwidth}
		\begin{threeparttable}[]
			\begin{tabular}{lll}
				\toprule
				\textbf{Paper} & \textbf{Related Keyword} &
                \textbf{Publish Date} \\
				\midrule
                Database Computers? A Step Towards Data
                Utilities \cite{database-computer} & Database
                Computer & xx-12-1976 \\
				Active disks: programming model, algorithms and
                evaluation \cite{active-disk-pillar}\footnotemark[3] &
                Active Disks & xx-10-1998 \\
                The Necessary Death of the Block Device
                Interface \cite{death-interface} & Host-managed & xx-06-2012 \\
                Active Flash: Out-of-core data analytics on flash
                storage \cite{active-flash-piller} & Active Flash & 16-04-2012 \\
                Near-Data Processing: Insights from a MICRO-46
                Workshop \cite{2014-micro-ndp} & Near-Data Processing (NDP) &
                06-08-2014 \\
                Computational Storage: Where Are We
                Today? \cite{barbalacecomputational} & Computational Storage &
                11-01-2021 \\
                Computational Storage Architecture and Programming Model
                v0.8 \cite{snia-model} & Computational Storage & xx-06-2021 \\
				\bottomrule
			\end{tabular}
			\begin{tablenotes}[para,flushleft]
				\centering List of initial seed papers used to construct list of
                keywords.
			\end{tablenotes}
		\end{threeparttable}
	\end{adjustbox}
\end{table}

\footnotetext[3]{Several works around utilizing embedded processors on hard
disks for offloading computations have been introduced around the same time.
We have chosen the work on Active Disks as a technical report on the same
topic by the same authors was released 7 months prior.}

Prior to identifying keywords we first define the inclusion and exclusion
criteria in the next section.


\subsection{Inclusion \& Exclusion criteria}

This section defines the inclusion and exclusion criteria used throughout this
work. For a work to be considered for inclusions it must match all inclusion
and none of the exclusion criteria. Any notable exceptions to this will be
indicated explicitly.

\begin{itemize}
    \item I.1 - The work brings computations closer to the (non-volatile) 
          storage layer.
    \item I.2 - The proposed architecture must match the SNIA Computational
          Storage definition \cite{snia-model}.
    \item I.3 - The proposed solution must be novel.
    \item E.1 - The work involves Processing-in-Memory (PIM).
    \item E.2 - The work involves micro-architectural or computer-architectural
          changes to physically bring computational elements closer to memory
          (on die).
    \item E.3 - The work involves a literature survey.
    \item E.4 - The work is a continuation or improvement of a previous work.
\end{itemize}

These criteria aim at extracting only the most notable works reducing data
movement or offloading host computations with a limited selection. The
inclusion criteria should allow to achieve this. In addition we want to be able
to identify fundamental characteristics as well as limit the scope so all works
have similar challenges. Similarly, the exclusion criteria are designed to
achieve this.

As mentioned we allow for a limited selection of 50 papers of which 25
dedicated to PFS. That does not mean that the entire list of references will be
limited to these 50 papers. Rather, only these works are used for answering our
research questions. Any support references to further amplify our arguments or
better illustrate our case will still be included.

Apart from this selection methodology we have one notable exception that clearly
breaks our criteria. This is the work by Antonio Barbalace \&
Jaeyoung Do \cite{barbalacecomputational}. This work provides considerable
contribution that strongly overlaps with our
\textit{What are the challenges involving PFS} research question. Given this
overlap it would be unfair to exclude this work.

With these criteria we can now perform some initial unstructured exploration
using the seed papers to determine a list of relevant keywords. These keywords
are described in the next section.

\subsection{Exploratory Keywords}

As mentioned the keywords are extracted from seed papers and their citations.
The list of keywords is by no means exhaustive and aims at using the most
relevant keywords while also allowing for variety. An example of such variety
would be works on fixed CS which are less common than those
on programmable CS. The key difference between fixed and programmable storage
is that in programmable storage the program running on the device can be
changed. This can be done to various \textit{degrees of programmability} which
will be described in a later section.

The keywords are shown in table \ref{table:keywords} also showing the amount of
works initially selected per keyword as well as the platform. Also shown are
several compound keywords showing substitutions such as
\textit{User-Programmable (KVS / Storage)}. This means that we search for both
\textit{User-Programmable KVS} and \textit{User-Programmable Storage}. Finally
any characters commonly used to substitute spaces such as \textit{-} are used
in searches both as is and substituted with space.

For each keyword we limit the number of evaluated works to about 10 on each
individual platform. With the exception for the \textit{Near-data processing},
and \textit{Computational storage} keywords. The selection of works considered
for evaluation is based on the content of the abstract and references. 

Several libraries also include work of other publishers. Such is the case
with the ACM library. In those cases we count the work towards the publisher not
the maintainer of the library. However, should the platform itself not be a
publisher as with Semantic scholar Then it will still count towards Semantic
Scholar. Our approach is not used to conclude anything about the distribution of
CS works across platforms. Rather, it is used to prevent missing any relevant
works.

In total 163 different works are selected across all platforms and keywords.
The distribution of these works is shown in table \ref{table:keywords}. However,
the summation of these totals does not result in 167, the reasoning behind this
is explained in the following reproduceability section.

\begin{table}[h!]
	\caption{Initial Keyword Selection Overview}
	\label{table:keywords}
	\centering
	\begin{adjustbox}{width=1\textwidth}
		\begin{threeparttable}[]
			\begin{tabular}{lllllll}
				\toprule
				\textbf{Keyword} & \textbf{ACM} & \textbf{IEEE} &
                \textbf{SNIA} & \textbf{Usenix} & \textbf{Semantic Scholar} &
                \textbf{Total} \\
				\midrule
                Active Disks & 5 & 2 & 0 & 1 & 9 & 17 \\
                Active Flash & 5 & 4 & 0 & 4 & 2 & 15 \\
                Active Storage  & 4 & 5 & 0 & 3 & 7 & 19 \\
                Intelligent Disk & 1 & 0 & 0 & 0 & 1 & 2  \\
                Near-data Processing & 10 & 6 & 2 & 0 & 2 & 20 \\
                Computational Storage & 7 & 0 & 18 & 1 & 7 & 33 \\
                in-storage (computing / computation) & 5 & 10 & 4 & 0 & 4 & 23 \\
                (User-)Programmable (KVS / Storage) & 2 & 1 & 0 & 1 & 2 & 6 \\
                Smart SSD & 7 & 2 & 3 & 0 & 0 & 12 \\
                Storage Computing & 2 & 1 & 0 & 1 & 0 & 4 \\
				\bottomrule
                Totals & 48 & 31 & 27 & 10 & 34 & 150 \\
			\end{tabular}
			\begin{tablenotes}[para,flushleft]
				\centering Overview of initially selected works across keywords
                    used in literature study.
			\end{tablenotes}
		\end{threeparttable}
	\end{adjustbox}
\end{table}

\subsubsection{Selected Literature}

From these initial works the exclusions and inclusion criteria are used to
create a limited set of around 50 works. The distribution across keywords and
platforms for this limited selection is shown in table
\ref{table:keywordsselection}. For each keyword and platform we also show how
many works apply as PFS.

\begin{table}[h!]
	\caption{Filtered Keyword Selection Overview}
	\label{table:keywordsselection}
	\centering
	\begin{adjustbox}{width=1\textwidth}
		\begin{threeparttable}[]
			\begin{tabular}{lllllll}
				\toprule
				\textbf{Keyword} & \textbf{ACM} & \textbf{IEEE} &
                \textbf{SNIA} & \textbf{Usenix} & \textbf{Semantic Scholar} &
                \textbf{Total} \\
				\midrule
                Active Disks & 3 (1) & 1 (0) & 0 & 0 & 1 (0) & 5 (1) \\
                Active Flash & 0 & 2 (2) & 0 & 1 (1) & 0 & 3 (3) \\
                Active Storage  & 1 (0) & 0 & 0 & 1 (0) & 2 (0) & 4 (0) \\
                Intelligent Disk & 1 (0) & 0 & 0 & 0 & 0 & 1 (0) \\
                Near-data Processing & 7 (5) & 1 (0) & 0 & 0 & 0 & 8 (5) \\
                Computational Storage & 2 (2) & 0 & 7 (0) & 0 & 3 (2) & 12 (4) \\
                In-storage (Computing / Computation) & 4 (3) & 1 (1) & 2 (0) & 1 (0) & 2 (2) & 10 (6) \\
                (User-)programmable (KVS / Storage) & 2 (2) & 0 & 0 & 1 (1) & 3 (0) & 6 (3) \\
                Smart SSD & 2 (2) & 1 (1) & 1 (1) & 0 & 0 & 4 (4) \\
                Storage Computing & 2 (1) & 1 (1) & 0 & 0 & 0 & 3 (2) \\
				\bottomrule
                Totals & 28 (16) & 7 (5) & 10 (1) & 4 (2) & 11 (4) & 60 (28) \\
			\end{tabular}
			\begin{tablenotes}[para,flushleft]
				\centering Overview of selected works across keywords in
                    literature study. The number of works categorized as PFS are
                    shown using the round brackets.
			\end{tablenotes}
		\end{threeparttable}
	\end{adjustbox}
\end{table}

\subsection{Literature Analysis}

The thorough evaluation of works allows for comprehensive insight in to how
particular keywords are used. We share this disambiguation here as it can
greatly improve readability, especially for those less familiar with the field
of CS. Together we show a visual timeline of when works where published and to
which keyword they relate as well as to what degree they are incorporated in
this work.

\subsubsection{Active Disks}

Starting with \textit{Active Disks} we see two uses for this keyword. First is
the initial use around 1998 till 2002 here we see the physical combination of a
compute element and a HDD. Second, is used to indicate the revival of this
paradigm with SSDs. Similarly in nature to the first use is the use of the
\textit{Intelligent Disk} keyword all be it less popular.

\subsubsection{Active Flash}

This paradigm of combining computational elements continues with
\textit{Active Flash} but naturally this keyword is focused on combining compute
With SSDs. In addition this is also how the \textit{Smart SSD} keyword is used.
We observe the first use of the \textit{Active Flash} keyword around 2011 well
into 2013 after which it seems to have become increasingly less popular. While
the use of \textit{Smart SSD} is observed later starting around 2013.

\subsubsection{Active Storage}

For \textit{Active Storage} we found works on scheduling task parallelism for
FPGAs \cite{10.1145/1391962.1391969} as well as general distributed object
storage \cite{380412}. This is at least partly due to inexperience of the reader
as none of these works actually contain the keyword \textit{Active Storage} in
the title or abstract. We will address further discrepancies like these in the
reproduceability section. The term is perhaps best described by Ilia Petrov, et
al in their survey on \textit{Active Storage}:
\begin{pquotation}{Ilia Petrov, et al, 2018 \cite{Petrov2018}}
Active Storage refers to an architectural hardware and software paradigm, based
on colocation storage and compute units. Ideally, it will allow to execute
application-defined data- or compute-intensive operations in situ.
\end{pquotation} With this definition in mind it is clear the keyword is
relevant to find related works. Although we more typically see
\textit{Active Storage} used in networked, distributed, HPC and big-data
applications such as with MapReduce \cite{10.1145/1327452.1327492}.

\subsubsection{Near-data Processing}

\textit{Near-data Processing} \cite{2014-micro-ndp} is used for both storage and
memory systems. Such as processors in memory channels or on 3D stacked memory
dies. Additionally, this term is also used in micro-architectural designs where
techniques are employed to bring the computational element, such as an
Arithmetic Logic Unit (ALU), closer to (memory) storage. Nevertheless, this
keyword is often used in influential CS works so all be it more labor intensive
it can not be omitted.

\subsubsection{Computational Storage}

With regards to other keywords, \textit{Computational Storage} (CS) is a
relatively new term. During our exploration the earliest occurrence was found in
2017 with the fast majority starting from 2019. The large amounts of work found
for this keyword is mainly due to the many presentations given at
SNIA SDC \cite{snia-sdc-conference} or the more recent SNIA PM+CS
Summit \cite{snia-pmcs-summit}. Over 15 of the works from the initial selection
on CS came from SNIA with 13, the vast majority, published in 2020 or 2021. It
is unclear however, if we will see this momentum for CS extend outside of SNIA.

\subsubsection{In-storage (Computing / Computation)}

With \textit{In-storage (Computing / Computation)} we are likely to find many
works related to PFS. We find the first works back in
2012 \cite{10.1145/2364489.2364497} although the use of this term in literature
remains sporadic until 2016 from which point on we see significant increase in
its usage \cite{10.1145/2933349.2933353, 7524716, 10.14778/2994509.2994512}.

\subsubsection{(User-)programmable (KVS / Storage)}

The use of \textit{(User-)programmable (KVS / Storage)} throughout our
selected literature is a bit mixed. We see the expected use of end-user
programmable CS systems. However, we also see works appear in these searches
related to MySQl query offloading. Still, the majority of works selected from
this keyword is on CS that offers some degree of user programmability.

\subsubsection{Smart SSD}

With \textit{Smart SSD} we found a similar amount of supportive literature as
well as literature directly related to CS. With supportive literature we mean
technologies our case studies that aid the implementation of CS our show its
necessity. In a later section we will go into more detail about some areas of
supportive literature.

\subsubsection{Storage Computing}

Although very few results are found for \textit{Storage Computing} as keyword,
overall it seemed to be effective at finding CS works. The use of this keyword
relates mainly to \textit{Active Storage} as almost all works proposed an
architecture in a distributed setting.

\subsubsection{Visual Timeline}

\begin{figure}[H]
    \centering
    \includegraphics[width=1\textwidth]{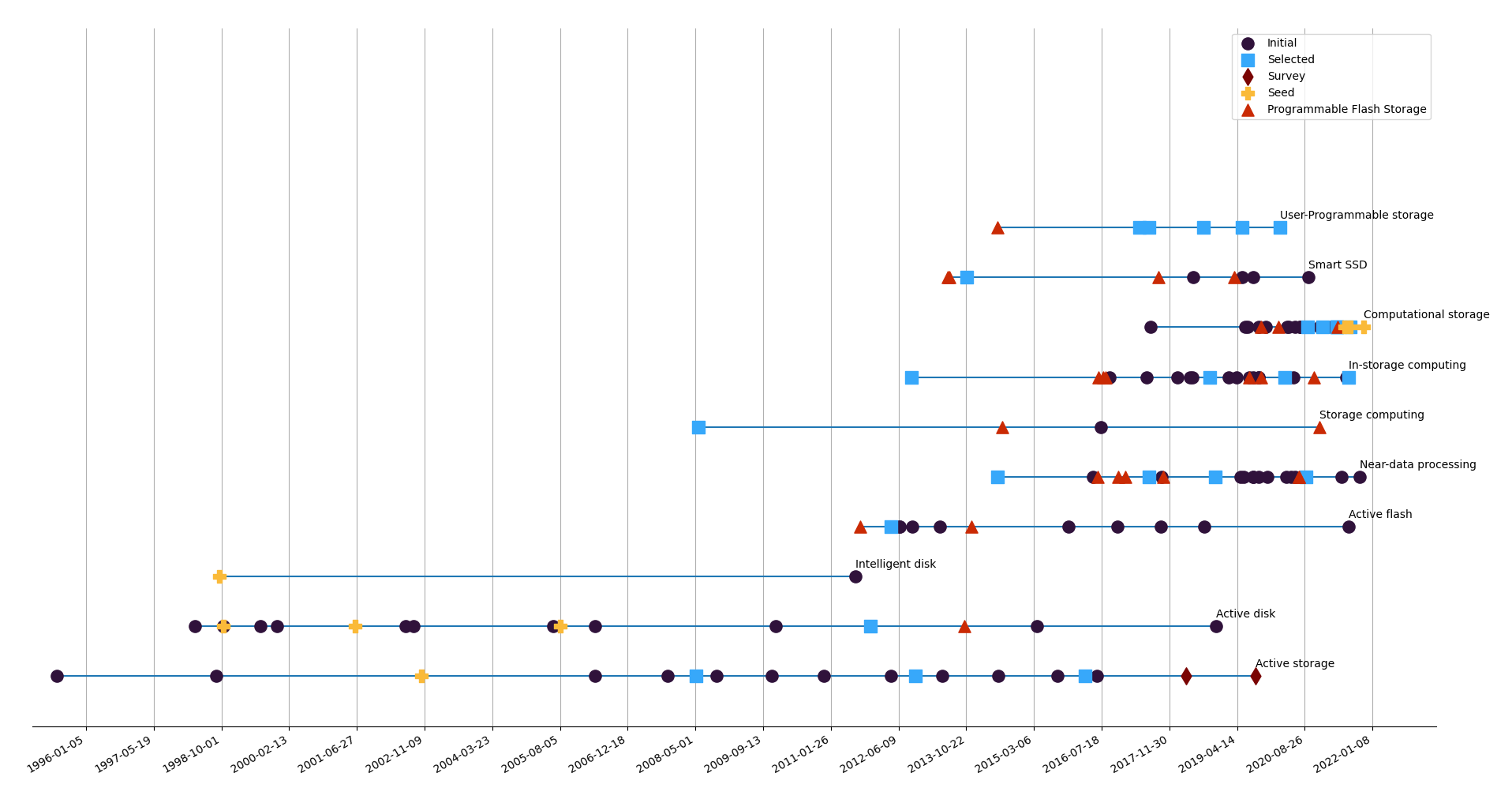}
    \caption{Overview of literature and their respective keywords projected
      on a timeline.}
    \label{figure:timeline-all}
\end{figure}

From these tables we derive a visual timeline that not only shows the initial
works found, selected or those attributed to PFS. In addition, it shows the
seed papers and surveys. This timeline is shown in figure
\ref{figure:timeline-all}. 

\subsection{Reproduceability}

The literature selection method introduces several activities that could hinder
reproduceability. In this section we describe these aspects and our way to go
about them.

\subsubsection{Search Order}

Firstly is platform and keyword search order. Finding related literature is
done by searching with keywords across several platforms. Naturally, some
platforms can return the same results as others. This will affect the
distribution as shown in tables \ref{table:keywords} and
\ref{table:keywordsselection}.

\noindent Our approach searches through the platforms in order of:

\begin{enumerate}
    \item ACM
    \item IEEE
    \item Usenix
    \item SNIA
    \item Semantic Scholar
\end{enumerate}

\noindent While the keywords are searched in the same order as they appear in
the tables \ref{table:keywords} and \ref{table:keywordsselection}, being:

\begin{enumerate}
    \item Active Disks
    \item Active Flash
    \item Active Storage
    \item Intelligent Disk
    \item Near-data Processing
    \item (User-)Programmable (KVS/Storage)
    \item Smart SSD
    \item Storage Computing
\end{enumerate}

\subsubsection{Keyword Attribution}

The methodology employed further influences how keywords are attributed to
literature beyond just the search order. This is because our methodology is
inherently naive. If a work is found using a given keyword it is always
attributed to this keyword no matter the content of the title or abstract. In
hindsight, a better approach would be to determine the matching keyword based
on the number of occurrences of keywords in the title and abstract, only resorting
to attributing it to the search query if no match is found.

\subsubsection{Novelty Assessment}

One of the selection criteria requires the work to be novel. Naturally there are
many different ways to assess novelty. Inherently, this can make the novelty
assessment quite subjective. To alleviate this we define a process for novelty
assessment that can easily be replicated by others. In this work novelty is
assessed based on the overlap between works released prior compared to the works
from our initial selection. Should a scientific contribution be released in 2013
then only works prior to this can invalidate its novelty. Overlap itself is
assessed by determining the proposed architecture, its user programming interface,
the underlying hardware abstraction layer (HAL) and its guarantees such as
multi-user tenancy or access level control (ACL).

\subsubsection{Support Files}

To orchestrate the collection of literature several support files are used.
These files are made available to further improve the reproduceability of this
work. The include a color coded overview of the evaluated literature, to which
platform it belongs and to which keyword. Other additional information includes
the DOI, date and if it regards PFS. These support files can be downloaded
online \cite{literature-support}.

\section{What is Computational Storage}

Computational storage is the broad concept of bringing computations closer to
the storage layer. It is the process of coupling Computational Storage
Functions (CSF) to storage, thereby, offloading the host or reducing data
movement.

\begin{pquotation}{SNIA, 2021 \cite{snia-model}}
    Computational Storage: Architectures that provide Computational Storage
    Functions coupled to storage, offloading host processing or reducing data
    movement.
\end{pquotation}

\begin{figure}[H]
    \centering
    \includegraphics[width=1\textwidth]{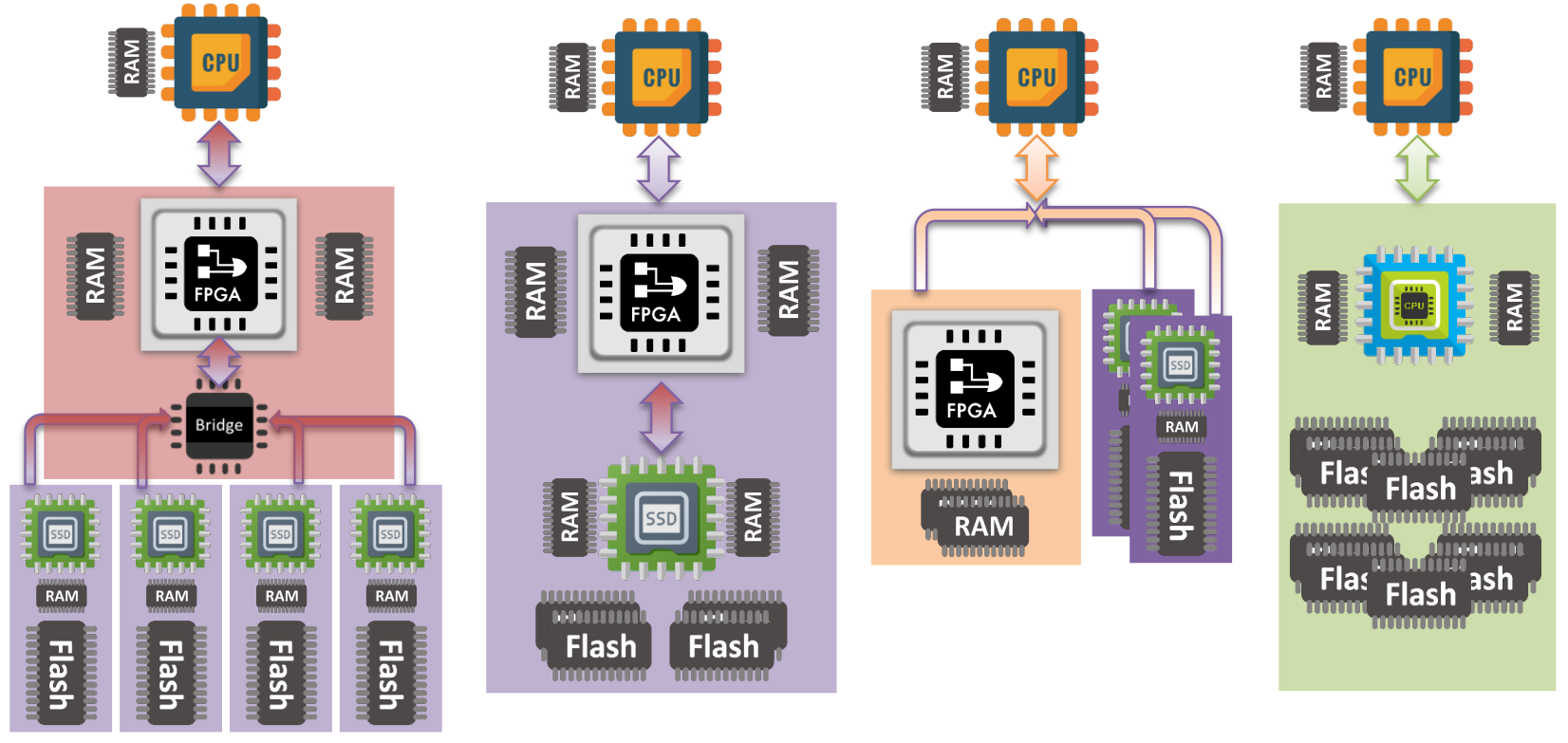}
    \caption{Overview of different CS architectures, image from Leah Schoeb \cite{schoeb_2020}}
    \label{figure:cs-architectures-visual}
\end{figure}

Given this flexibility CS sees numerous different architectures that can be
categorized into three predominant types as shown in
figure \ref{figure:cs-architectures-visual}. This figures uses examples showing
specfic types of computational elements and storage technologies while in
practice no restrictions apply here. Starting on the left we see an FPGA being
used as a central interface to communicate with multiple SSDs. All
communications from the host are done through the interface offered by the FPGA.
This architecture is of an array type. Moving one to the right is an FPGA used
to interface with a single SSD. This is known as drive type. When the FPGA is
replaced with a more conventional CPU it is also known as a drive type. When
the compute element and the SSD share the same bus as the host it is known as a
processor type. This type is illustrated from the right half of the center.

\begin{figure}[h]
    \centering
    \includegraphics[width=1\textwidth]{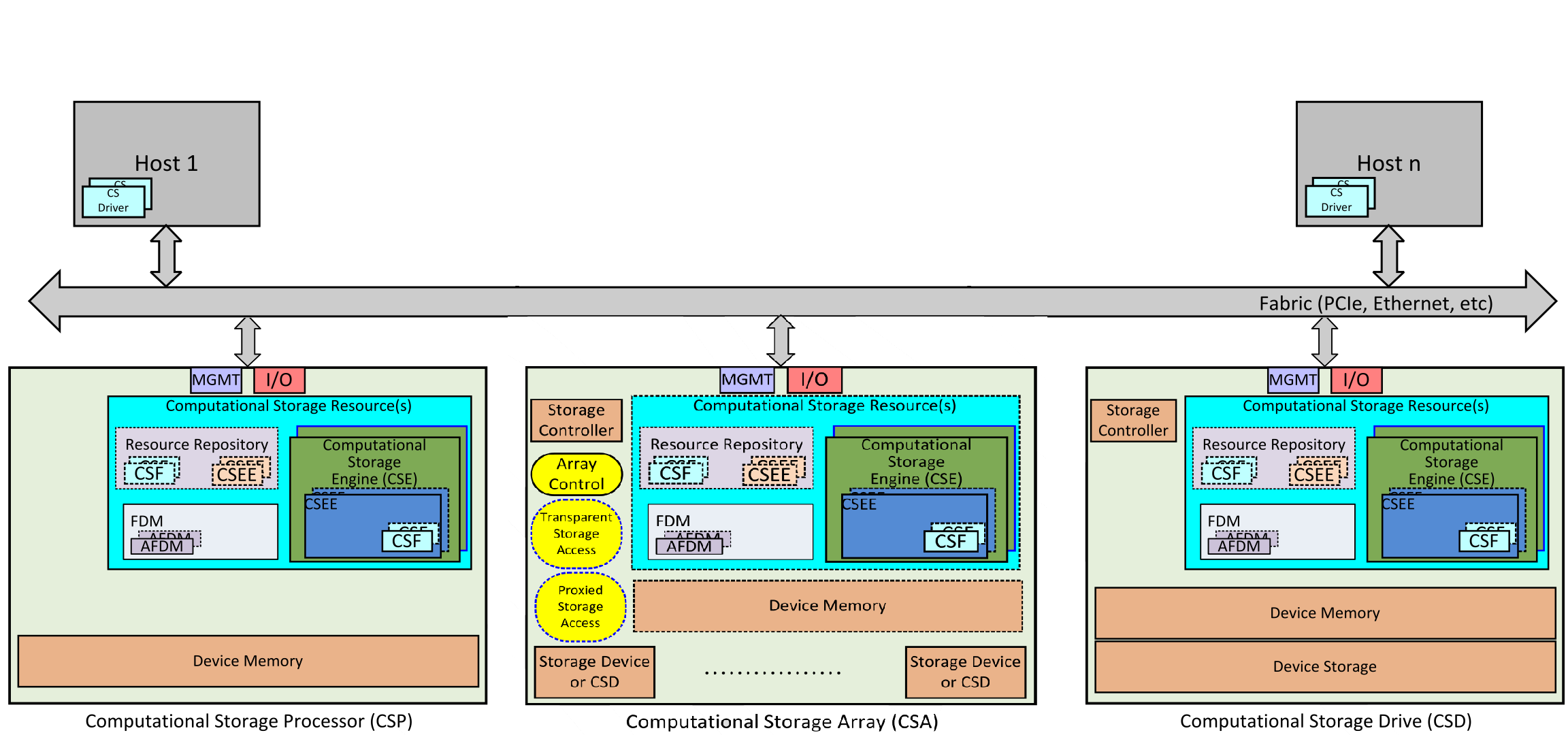}
    \caption{Overview of different CS architectures using more formal
        definitions and highlighting composition.}
    \label{figure:cs-architectures-schematic}
\end{figure}

Having understood these basic type abstractions we can go into the more formal
definitions as set out by SNIA. Here we see Computational Storage Processors
(CSP), Computational Storage Arrays (CSA) and Computational Storage Drives
(CSD). The more general term Computational Storage Device (CSx) applies to all
three categories. In figure \ref{figure:cs-architectures-schematic} we show
these architectures and their individual components.

Out of these components only a few are essential to build a CSx. These include
a Computational Storage Engine (CSE),
Computational Storage Execution Environment (CSEE),
Computational Storage Function (CSF), Computational Storage Resource (CSR),
Function Data Memory (FDM) and Allocated Function Data Memory (AFDM).

The overall idea is that a CSR has one or more CSEs that contain one or more
CSEEs that subsequently contain one or more CSFs. The CSR can allocate FDM to
create AFDM used by a CSF during its execution inside a CSEE.

The CSR is the overall part of the device, accessible over an interface, that
contains all the components to support CS. Inside the CSE is the computational
element, this can be a Field Programmable Gate Array (FPGA), CPU or Digital
Signal Processor (DSP). To run CSFs on these CSEs we need an CSEE. These are
essentially support libraries, sandboxing or virtualization layers.

Using this basic architecture would create a fixed CSx that has a predefined set
of CSFs that can be executed and that can not be changed by the end user.
Contrarily, should the CSx contain a resource repository and provide a method
to upload new CSFs or CSEEs to this repository than it is a programmable CSx.

This line can become blurry as some devices allow to program a chain or Directed
Acyclic Graph (DAG) of unchangeable preprogrammed functions that can arguably
be seen as some form of programmability. Using this method it might even be
possible to support dataflow programming models.

Lastly, CS should not be confused with Programming-in-Memory (PIM) or smart
Network Interface Controllers (NIC). Particularly PIM might be misinterpreted
as CS since we have seen many recent works introduce PIM to non-volatile RAM
technologies. Similarly, although smart NICs should not be considered CS we see
many distributed CS architectures emerge that use a networking interface and
should be considered CS. The key difference here is that a smart NIC does not
contain any storage technology on the device itself, contrarily to CSxs with
a networking interface.

In short CS is the process of coupling any computation to storage thereby
offloading host processing or reducing data movement.

\section{History of Computational Storage}

Although the term \textit{Computational Storage} itself is relatively new, the
history of CS can be traced back to database computers developed in the 1970s
and 1980s \cite{database-computer}. These machines typically offloaded specific
database operations although this was often complicated by the infancy of
relational databases themselves as well as lack of associative memory.

Overall the entire history of CS is much richer than will be shown in this
section. This work has keeps this brief, focusing on the most prominent
contributions of the time. This is done to be able to achieve a more in depth
analysis on the past decade of CS works. An overview of the works covered in
this section is shown in table \ref{table:historycs}.


\begin{table}[H]
	\caption{Historical Computational Storage}
	\label{table:historycs}
	\centering
	\begin{adjustbox}{width=1\textwidth}
		\begin{threeparttable}[]
			\begin{tabular}{llll}
				\toprule
				\textbf{} & \textbf{Disc Architecture} &
                \textbf{SNIA Architecture} & \textbf{Release Date} \\
				\midrule
                CASSM \cite{10.1145/1282480.1282518} & PPT & CSD & 1975 \\
                RARES \cite{10.1145/320434.320447} & PPT & CSD & 1976 \\
                RAP \cite{10.1145/320544.320553} & PPT & CSD & 1977 \\
                DIRECT \cite{10.1145/582095.582098} & Per device & CSP & 1979 \\
                Active Disks, Riedel et al \cite{1558605665} & Per device & CSD & 1998 \\
                Active Disks, Acharya et al \cite{active-disk-pillar} & Per device & CSD & 1998 \\
                Intelligent Disk \cite{intelligent-disk} & Per device & CSD & 1998 \\
				\bottomrule
			\end{tabular}
			\begin{tablenotes}[para,flushleft]
				\centering Overview of historical CS works with their
                architecture and release dates.
			\end{tablenotes}
		\end{threeparttable}
	\end{adjustbox}
\end{table}

\subsection{Database Computers}

\begin{figure}[h]
    \centering
    \includegraphics[width=1\textwidth]{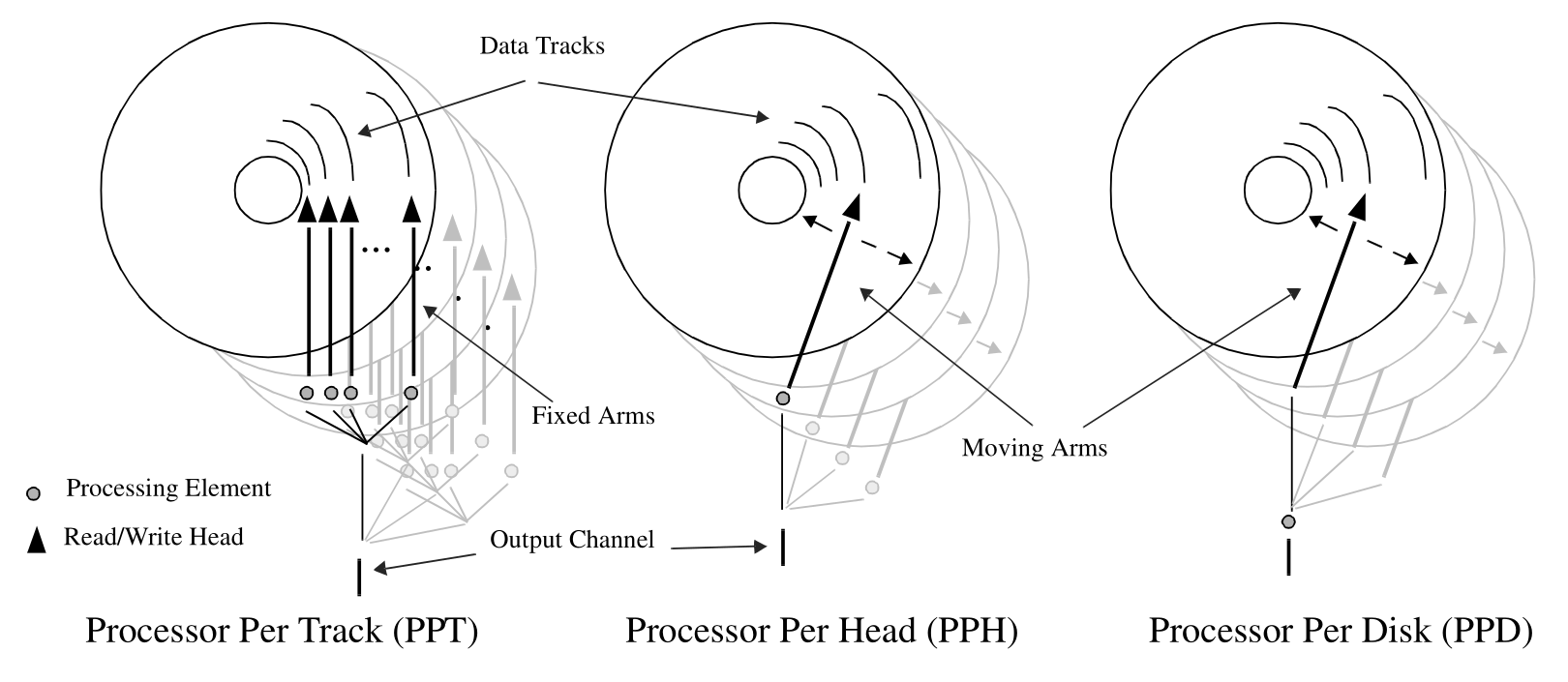}
    \caption{The three fundamental types of database computers employing
        different amounts of processing per harddrive component, image from
        \cite{active-disks-thesis}.}
    \label{figure:database-architectures}
\end{figure}

Several database computers employed an architecture to solve the issues of
their requirements directly such as
\textit{Content-Addressable Segment Sequential Memory}
(CASSM) \cite{10.1145/1282480.1282518}. This database computer used a
head-per-track disc and a Processor Per Track (PPT) architecture. In addition
Processor Per Head (PPH) and Processor Per Disc (PPD) architectures existed.
The difference between a PPT and PPH architecture in a head-per-track disc is
that in PPT the heads are fixed and in PPH the head is moveable. There are three
fundamental types of database compute architectures as shown in
figure \ref{figure:database-architectures}.

This terminology of PPT, PPH or PPD is sometimes also referred to as Logic Per
Track (LPT), Logic Per Head (LPH) or Logic Per Disc (LPD). These different types
of database computer architectures best relate to the CSD architecture from
SNIA. The CSA architecture does not really apply as it would be hard to argue that
individual tracks or heads on a disc are entire storage devices by themselves.

After CASSM we saw the introduction of \textit{Relational Associative Processor}
(RAP) \cite{10.1145/320544.320553}. This allowed for more complex searches
through metadata mark bits. In addition, this metadata was no longer stored in
RAM as was the case with CASSM but alongside the data itself. Similar to CASSM,
RAP used an PPT architecture.

Another machine was the \textit{Rotating Associative Relational Store}
(RARES) \cite{10.1145/320434.320447}. RARES used a PPT architecture as it
used a head-per-track disc with fixed heads. RARES is different in the sense
that it separates the processing into back-end and front-end processing as well
as storing data orthogonality to the storage layout. That is to say the data is
spread across tracks rather than being written linearly on a track.

Later database computers often did away with the PPT or PPH architectures such
as with DIRECT \cite{10.1145/582095.582098}. Here a crossbar switch was used to
allow any processor to communicate with any storage unit. In addition it used
a Multiple Instruction, Multiple Data (MIMD) instruction set as opposed to
Single Instruction, Multiple Data (SIMD) instruction set. This is one of the few
database computers to not use a CSD like architecture but instead use a CSP
architecture.

Even later machines did away entirely with custom built processors and started
using general purpose processors. Some commercially available systems, mostly
using general purpose processors are listed below.

\begin{itemize}
    \item Content-Addressable File Store (CAFS and SCAFS)
    \item Microsoft TerraServer
    \item Digital TPC-C
    \item Teradata optical storage processor
\end{itemize}

The complexity that arouse from using custom processors with custom microcode
often made programming database computers difficult. Their performance also
typically did not outweigh their large size and cost. Even though the second
generation of database computers used commodity hardware. This still led to most
database computers not being attractive for practical applications outside of
research areas. Especially once networking, memory and processing power
become more powerful and cheaper, the need for these high cost purpose built
machines decreased. A noteworthy exception is the Teradata optical storage
processor that was still being sold at the end of the
1990s \cite{Ramsay1990IntegrationOT}.

\subsection{Active Disks}

From the research of database computers spawned the field of
\textit{Active Disks} in the late 1990s. These similarly, introduced processing
capabilities on the HDD itself but using a PPD or processor per device
architecture. Like database computers these are also CSD architectures.
Different from database computers in the sense that a single Active Disk could
be installed in a consumer computer or datacentre server. The necessity of a
large dedicated rack for a special purpose machine was thereby removed. This
also significantly reduced the cost as the additional processing would only be
around 10\% of the total cost of a HDD at that time. Although it did push the
entire responsibility of interfacing and programming these \textit{Active Disks}
to the host operating system.

\begin{figure}[h]
    \centering
    \includegraphics[width=0.5\textwidth]{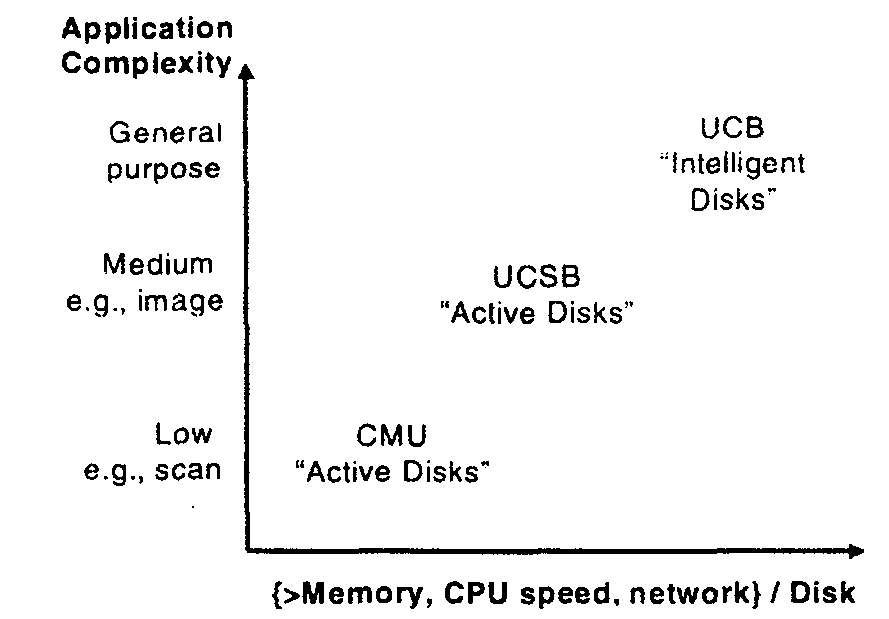}
    \caption{Different types of Active Disks technologies and their
        respective degrees of programmability, figure
        by Keeton et al \cite{intelligent-disk}.}
    \label{figure:active-disk-types}
\end{figure}

During the start of Active Disks as a research field we saw three primary
contributions. Two named after Active Disks \cite{1558605665, active-disk-pillar}
and one named Intelligent Disk \cite{intelligent-disk}. All three they offered
different architectures and degrees of programmability\footnotemark[4] as shown
in figure \ref{figure:active-disk-types}.


\footnotetext[4]{We will address \textit{degree of programmability} with much
more detail in sections on PFS.}

In increasing degrees of programmability we first start with Active Disks by
Riedel et al (CMU) \cite{1558605665}. These used a PPD architecture and used
scan like operations to reduce data transfers. While the work by Acharya et
al (UCSB) \cite{active-disk-pillar}. used a processor per device architecture.
A key difference is that a single device can contain multiple discs so PPD
potentially has more processors than a processor per device architecture.
Their work is also more focused on programming models, particularly streaming.
Their methodology expressed operations in so-called disklets.

Finally, Intelligent Disk \cite{intelligent-disk} used an architecture
similar to that of DIRECT \cite{10.1145/582095.582098} using a crossbar network
of fully connected HDDs. The key difference here is that each HDD has its own
dedicated processor and memory while on DIRECT these where separated. While
the programmability in intelligent disk is flexible it provides no programming
model itself beyond a few basic recommendations. In addition the work does not
provide any experimental setup or evaluations unlike the two other works on
active disks.

Beyond this initial surge of new interest in CS research related to active
disks, the field soon stagnated. There was some work mainly on applications in
Redundant Array of Independent Disks (RAID) or Networked Attached Storage (NAS)
deployments but those are considered outside the scope of this work.

\subsubsection{Active Disks Lack of Success}

A number of reasons might have contributed to the lack of success for active
disks. One prominent one is the cumbersome programming and development
methodology. Especially in a time where OpenMP (2000) and OpenMPI (2004) started
to emerge. These libraries significantly reduced the development effort
required for multi-threaded and distributed processing. Although it took until
around 2011 until we saw support for OpenMPI in HPC batch processing systems
like SLURM\footnotemark[5]. The availability of multi-threaded or distributed
processing is a direct competition for the case of CS. Since these additional
computational capabilities can resolve computational bottlenecks just active
disks could.

\footnotetext[5]{For example, HTCondor still does not support OpenMPI and
perhaps never will.}

These initial development methodologies further complicated their use as they
poorly addressed multi-user tenancy, security and safety guarantees such as
concurrent access and modification.

Beyond the cumbersome software development we saw the cost of HDDs drop over
the course of the early 2000s, their capacity increase rapidly while there
number of I/O operations per second (IOPS) stagnated. Meanwhile the performance
of desktop and server grade processors increased, particularly in clock speed
going from 600 MHz in 2000 to 2.8 GHz in 2010. This meant processors could serve
much more interrupt requests at any given time as the amount of instructions it
takes to serve an interrupt service routine has remained relatively the same.
The stagnated IOPS for HDDs combined with faster interrupt processing meant that
it was impossible for a modern CPU to be overloaded by one or even a large
amount of harddrives. Deployments of 64 or 128 HDDs in a single server where not
uncommon around that time.

The peripheral interconnects also saw advancements such as with the
introduction of SATA as opposed to IDE. As well as the introduction of AHCI.
This simplified the communication with HDDs from the point of the host processor
as well as offering significantly higher bandwidth.

\section{The Ubiquity of Computational Storage}

As seen with the history of CS, the need to reduce data movement and thus
move computations to storage has been prevalent for quite some time. In this
section we will showcase how many different fields are already actively using
this technique to advance their field. Although often not directly attributed
CS, it is clear that in these fields we see architectures similar to those of
CS. Even more so, should we finally see readily available PFS\footnotemark[6],
than it is extremely likely that these fields would start using them as well.

Before describing all the different fields we CS emerge we briefly look at
distributed storage architectures.

\footnotetext[6]{With a stable interface and architecture that is not prohibitively
expensive or difficult to use..}

\subsection{Layered Distributed Storage Architectures}

\begin{figure}[h]
    \centering
    \includegraphics[width=0.8\textwidth]{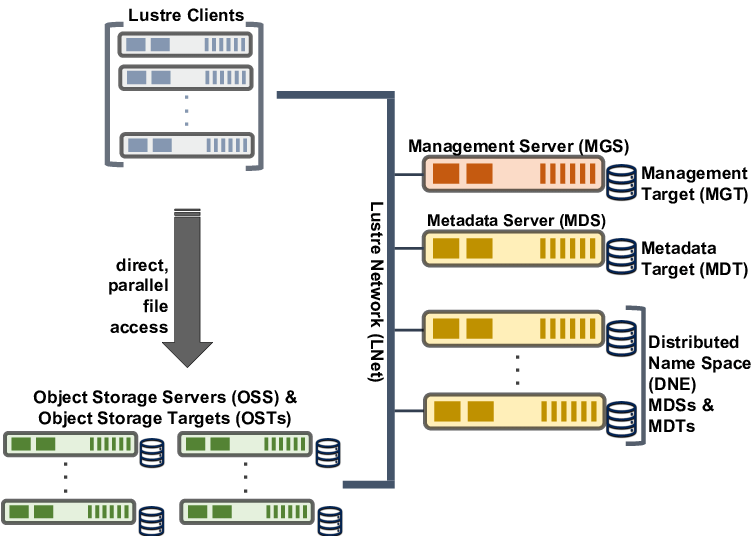}
    \caption{Overall architecture of Lustre parallel distributed filesystem,
        figure by Arnab et al \cite{10.1109/CLUSTER.2019.8891045}.}
    \label{figure:distributed-fs-arch}
\end{figure}

One prominent application for CS is in distributed filesystems. These
filesystems are often comprised of several fundamental components. Typically
explicitly separated into subcomponents but sometimes architectured as a
monolith. The typical subcomponents of a distributed filesystem include a
key-value store or an object store. Sometimes an object store is a key-value
store but almost never the other way around. This object store is than used to
support the distributed filesystem. Other components such MetaData Servers (MDS)
exist in some architectures although this is sometimes stored alongside the
objects in the object store.  Lastly, When an entire storage device is claimed
for use as an object store it is referred to as an object-based storage device
(OSD). In figure \ref{figure:distributed-fs-arch} an example of a distributed
filesystem architecture is shown, Lustre in particular.

Other configurations are also possible such as building a local filesystem on
a distributed object store. Historically the key-value and or object store where
written on top of a traditional filesystem but strong evidence suggest that
this can severely degrade performance \cite{10.1145/3341301.3359656}. Lastly,
special purpose HDDs exist that run object stores with complimentary API, an
example is Seagate their Kinetic HDD \cite{seagate_kinetic}.

CS is applied in all levels of distributed filesystems to various degrees. The
understanding of these basic levels is thus essential in identifying how key CS
works are being applied here. In the field of \textit{Active Storage} we often
look at the object store or the filesystem as a whole. While sometimes
literature on active storage solely looks at the object store. The term active
storage is typically not used in the context of key-value stores.

\subsection{Active Storage}

Research in the field of \textit{Active Storage} has allowed object stores and
distributed filesystems to better leverage compute capabilities of storage
nodes. Starting with the introduction of scriptable Remote Procedure Calls (RPC)
in 2002 \cite{10.1145/605432.605425}. sRPC allows to offload computations by
transmitting tcl scripts over the network subsequently executed by the
receiver. Similarly, Splinter allows to export Rust scripts over the network
that can be remotely invoked while also using RPC \cite{222599}. Another
approach decomposes the computation into predefined steps called
\textit{functors} \cite{Wickremesinghe02distributedcomputing}, effectively
allowing for programmability with a fixed function dataflow model. An
intermediate approach between these two allows tying user defined functions to
specific objects \cite{XIE2016746}.

Beyond entirely academic applications we also see active storage being applied
to existing distributed filesystems \cite{10.1145/1362622.1362660}. A key
benefit of this approach is that implementing important filesystem properties
such as multi-user tenancy, concurrent access and modifications, crash recovery
and self-healing might become substantially easier. This is because typically,
distributed filesystems already come with the majority of these features.

\begin{figure}[h]
    \centering
    \includegraphics[width=0.5\textwidth]{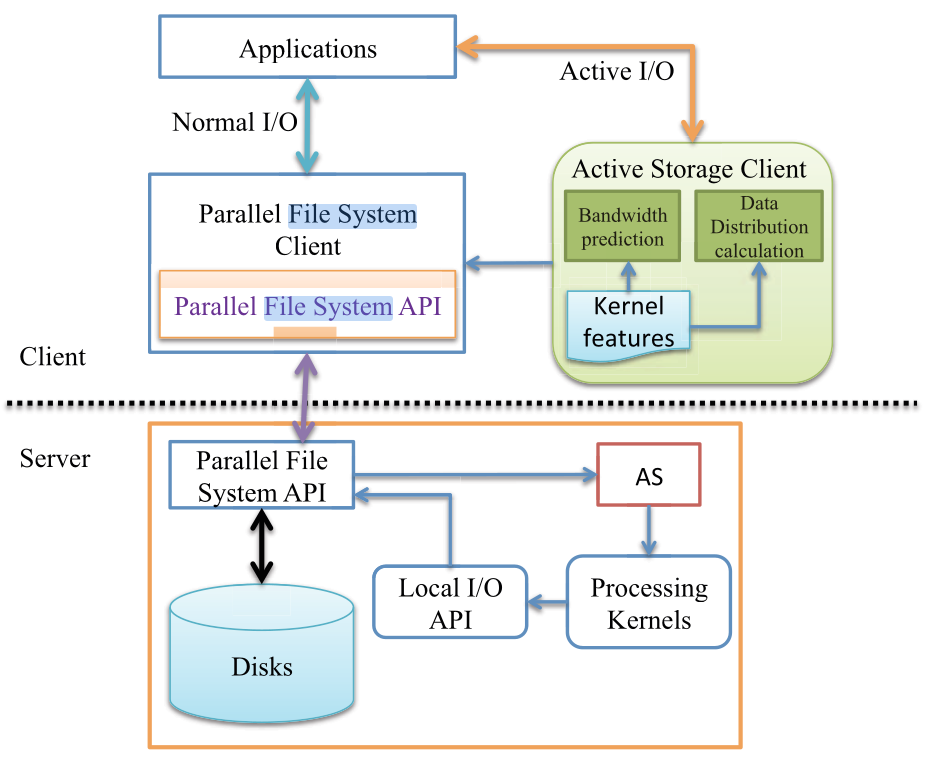}
    \caption{Active storage architecture supporting both regular access as well
        as offloaded, figure by Chen et al \cite{6337599}.}
    \label{figure:active-storage-arch}
\end{figure}

Not every application might be fit for active storage, that is why Chen et
al \cite{6337599} proposed a dynamic active storage system on top of Lustre. In this
work they reduced the overall resource contention and improved data dependency
management as well as allowing regular distributed filesystem access. We show
this high level architecture in figure \ref{figure:active-storage-arch}.

\subsection{Consumer Applications}

Outside of research the Hadoop distributed filesystem, HDFS, actively ensures
computations are being executed close to data. While RADOS, the object store
used by CephFS utilizes intelligence functions on storage
nodes \cite{10.1145/1374596.1374606}. Similarly, the CephFS Bluestore filesystem
supports transparent compressions for which the computations are done on the
local storage node \cite{10.1145/3341301.3359656}.

On the protocol side, NVMe has seen the addition of a copy command. Allowing to
copy data across SSDs without moving it to the host first \cite{9781424494323}.
Similarly, the use and applications for peer to peer dma over PCIe are
increasing as well \cite{amd_smart, lwn_p2pdma}, particularly due to NVMe-oF and
RDMA. With current trends of an increasing amount of special purpose processors,
perhaps it has become possible for an heterogeneous computing architecture of
autonomous special purpose processors to emerge \cite{10.1145/3458336.3465291}.

Clearly this mix of storage and computation is not to the same degree as in
active storage research. However, it might only be a matter of time before we
see more prominent CS applications appear across different fields.

A prime example of this new CS frontier is found in the I/O controller
of the Playstation 5. Here the use of transparent decompression, supporting zlib
or oodle kraken formats, can achieve between 8 and 9 GB/s transfer speeds on a
5.5 GB/s storage device\cite{ps5}.






\section{A Case for Programmable Flash Storage}

With the Von Neumann architecture we see an increasing amount of data movement
that is quickly becoming the bottleneck in many data driven
applications \cite{2016-western-digital}. Typically many of these data movements
are redundant wasting both time and power, reducing throughput and energy
efficiency.

Meanwhile, due to stagnated generational improvements in
CPUs \cite{cpu-improvements}, we are entering an age of specialization. current
generations of processors are being equipped with dedicated machine learning or
augmented reality accelerators. Key examples are the neural engines inside
Apples A14 and M1 processors. Researchers predict this trend of special purpose
hardware is only going to increase \cite{10.1145/3458336.3465291}.

These specialized accelerators can benefit from recent developments
in interconnects such CCIX \cite{ccix_interconnect},
GenZ \cite{genz_interconnect}, CXL2 \cite{cxl2_interconnect} and
OpenCAPI \cite{opencapi_interconnect}. Allowing for simpler or even cache
coherent communication in a heterogeneous architecture.

Similarly, we see the development of emerging open Instruction Set Architectures
(ISA) such as RISC-V \cite{riscv} and OpenPower \cite{openpower}. The key to open
standards like these is the ability for anyone to develop, test and iterate
prototypes without licensing costs. This greatly accelerates progress in
specific fields.

Even more so, we see progress in the field of open FPGA toolchains and
parameterized soft cores. This allows anyone to research custom CPU architectures
that might better fit these specialized applications.

Clearly, Programmable Flash Storage (PFS), user-programmable computational
capabilities placed right on the SSD itself, is an important next step in the
heterogeneous architecture of the future.

\section{A Decade of Programmable Flash Storage}

It has been near a decade of research in PFS \cite{6062973} and yet we see no
widespread adoption. Recently Antonio Barbalace et
al \cite{barbalacecomputational} described several of the reasons that are
preventing this adoption.

In this section we go over the past decade of literature and identify several
key aspects of each work. These pieces of literature are further categorized
into fixed function and programmable flash storage architectures. Moreover, We
will define each of these identified aspects being
\textit{Degree of programmability}, \textit{Programming model},
\textit{Interface} and
\textit{Computational Storage Execution Environment (CSEE)}. First, we define
some key software elements to integrate PFS architectures.

\subsection{Components of a Successful PFS Software Architecture}

\begin{figure}[H]
    \centering
    \includegraphics[width=1\textwidth]{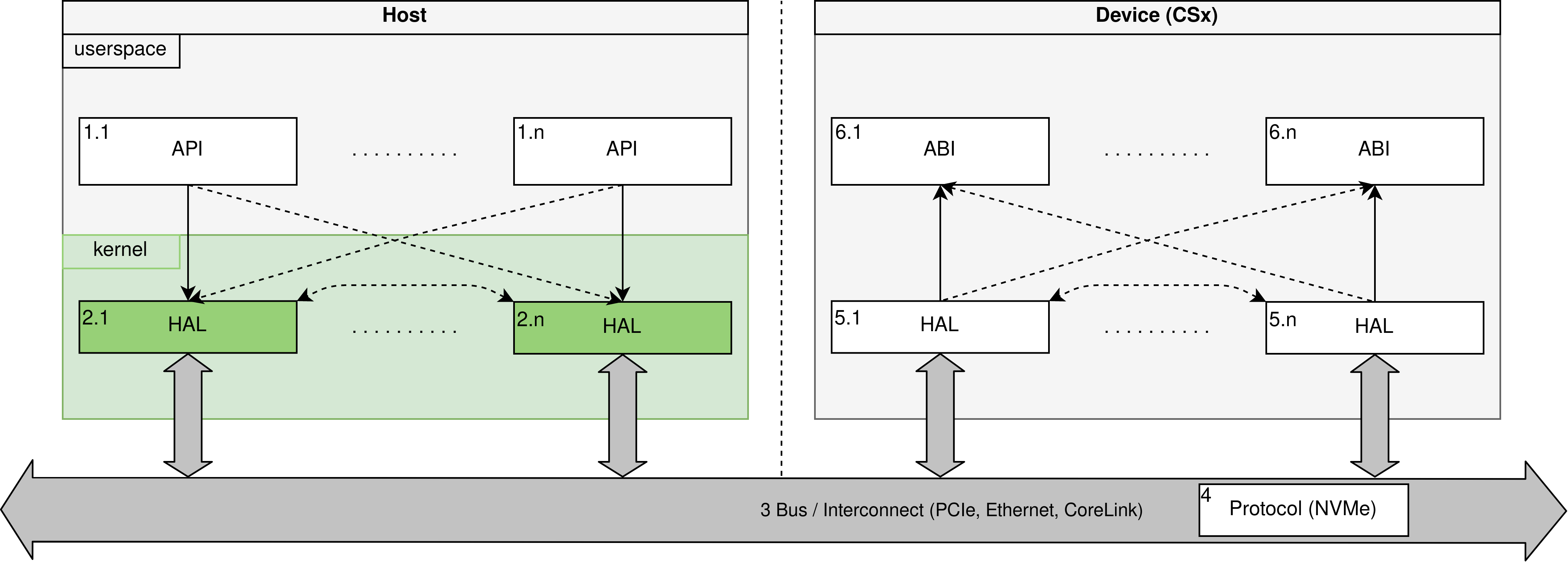}
    \caption{Overview of basic PFS software architecture allowing for
        full vertical integration.}
    \label{figure:basic-pfs-arch}
\end{figure}

A key to a successful PFS architecture is the complete design of a full vertical
integration. This is to say that implementing PFS requires making changes at
all layers of the software stack. Ranging from the user facing API
on the host down to hardware abstraction layers of the CSx we arrive at the
following components as shown in figure \ref{figure:basic-pfs-arch}.

First are the user facing APIs, supporting different programming models. These
define how a user has to write programs in order to interact with one or
multiple CSxs. A specific CSx could support one or more of these APIs with each
one or more programming models. Second is the one or multiple layers of
hardware abstraction (HAL), these typically interact with the kernel as well as
often being vendor specific. From the HAL the communication is managed over
buses and interconnects dictated by their communication protocols. Typical
examples for this third layer include PCI-e or CoreLink. Like the familiar
TCP/IP networking stack this third layer can contain one or more additional
specialized communication protocols, a typical example would be NVMe.
From here we arrive at the device with its own HAL layer. With the final sixth
layer being the ABI that user written programs must conform to.

It is the responsibility of both the host API and HAL layers to ensure
conformity to this ABI. While the device HAL verifies this. Similarly, it is
the host API that is responsible for converting the user written program, using
their programming model, into the ABI of the device.

This architecture is substantially different than the overview presented by
SNIA in figure \ref{figure:cs-architectures-schematic}. This is because we are
only looking at the essential components for full vertical integration required
for a programmable architecture. This architecture should be seen as
complementary to that of SNIA with further subdivisions being logical. However,
we have intentionally kept this figure relatively simply to more easily explain
the transnational interactions between different layers.

Typically the support for the third layer is largely performed in hardware with
the remainder carried out by the host operating system. Similarly, large parts
of the fourth layer are typically implemented by host operating systems.
Sometimes operating systems expose special drivers that allow to implement
the fourth or even the third layer in userspace. Examples of these drivers would
be uio\_pci\_generic \cite{pci_generic} and io\_uring \cite{io_uring} in Linux.

\subsubsection{Dynamic Device Discovery}

One key element is missing from our previous architecture however. This is
because tying many APIs to many HALs is cumbersome, error prone and discourages
vendor adoption, instead we propose a dynamic architecture utilizing
\textit{Installable Client Drivers (ICD)}. These ICDs are drivers discovered and
loaded at runtime, the interface is offered by the user facing API while they
are implemented by the vendor. The vendor is allowed to utilize any HAL or
additional components. Supporting multiple APIs or multiple devices with the
same HAL is also allowed. The only restriction is that the ICD must conform to
the interface of the loader which is dictated by the API.

\begin{figure}[H]
    \centering
    \includegraphics[width=0.8\textwidth]{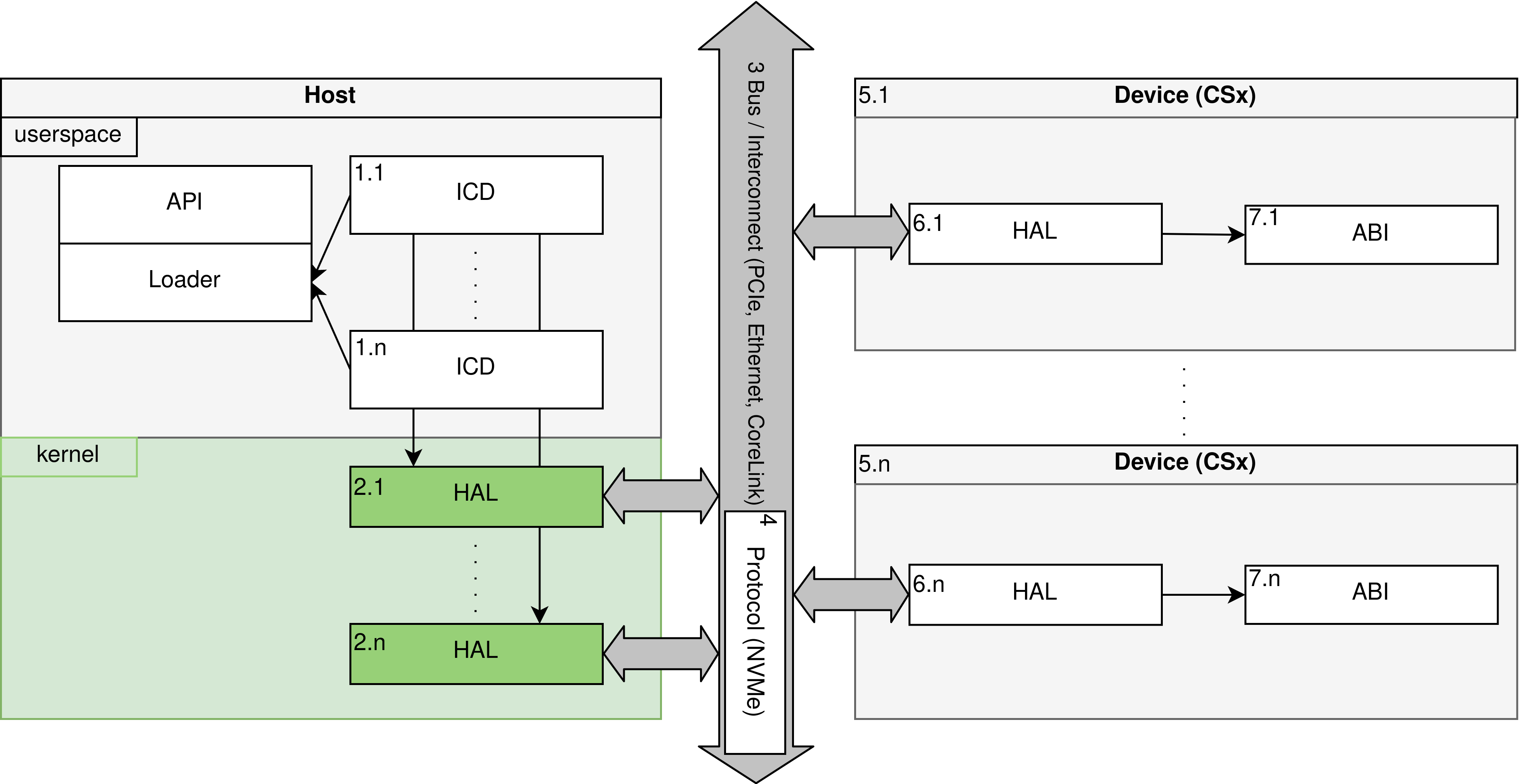}
    \caption{Overview of PFS loader architecture}
    \label{figure:loader-pfs-architecture}
\end{figure}

This approach is very similar to those of advanced computer graphics APIs such
as OpenGL and Vulkan. We show our mechanism for loading applications in
figure \ref{figure:loader-pfs-architecture}.

\subsubsection{We Present OpenCSL}

This overall presented system is called OpenCSL and will be released in a future
work. It will support all three major programming models as well as allowing for
agnostic vendor implementations. Moreover, the user written programs will be
portable across different devices of different vendors. Finally, OpenCSL will
address common problems of these systems such as multi-user tenancy and
security.

\subsection{Aspects of Programmable Flash Storage}

When identifying different aspects of PFS numerous software layers emerge. From
the perspective of the end user we identify each one using a top down approach.
First, is the end user API that is used to program or interface with the device.
This could be a server using HTTP/Rest or RPC. In the next tables we categorize
this as the \textit{programming model}. Next is the \textit{interface}, this is
the layer between the host and device often encapsulating additional interfaces.
Typical examples are SATA or PCIe with NVMe as a typically encapsulated
protocol.

Similar encapsulations are possible for the programming model where RPC calls
could encapsulate SQL queries, for example.

Beyond this interface we arrive at the device and its
\textit{Computational Storage Execution Environment (CSEE)}. This is the
environment in which the commands and programming of the end user are executed.
These range in increasing levels of software abstraction. From the perspective
of the CSEE we are only looking at the top level of the stack For instance,
running a container with CSFs often requires an underlying operating system but
will still be classified as \textit{container}.

The last layer is the \textit{degree of programmability} which we will describe
in the next section. Before this description we identify each of the possible
categories for each layer starting with the programming model.

In no particular order these are \textit{Dataflow (MapReduce, DAG)}, 
\textit{Client / Server (RPC, HTTP)}, \textit{Shared memory} and
\textit{Declarative (Regex, MySQL)}. Although simple categories, the correct
attribution of these categories can become quite complicated due to nuances.
For instance, if an end user needs to link to a new shared library to call
methods with MySQL queries as argument, what should the programming model be?
In this work we argue \textit{Shared memory} as the library is introduced in the
PFS work itself. If the library is part of preexisting software, SQLite for
example, it is categorized as \textit{Declarative}. This is because only the
declarative queries are interacting with the CSx. In these cases the existing
libraries are simply used as transparent interfaces to simply integration for
the end user.

We believe that the interfaces will be self explanatory enough. Meaning that
they won't need specific categories or further clarification. However, CSEEs do
and we identify the following categories:

\begin{enumerate}
    \item Bistream; bitstream programmed directly unto FPGA.
    \item Embedded; single program, single memory space, no OS just 'real mode'.
    \item Accelerators; OpenCL, Vulkan.
    \item Real-Time operating system;
    \item Operating System;
    \item Container;
    \item Virtual Machine;
\end{enumerate}

We believe most of these categories are self-explanatory. However, the naunce
between embedded platforms and accelerators has to be properly explained.
While embedded platforms might deal with simple static memory allocations we
expect accelerators to support implicit and explicit host-device memory
transfers. In addition we expect to have some control over device memory
allocation as well as requiring a scheduler to be present.

Finally, before describing the \textit{degrees of programmability} we will
elaborate on the difference between this aspect and the
\textit{programming model}. These might seem overlapping but they are used to
differentiate between the perspective of the device and end user. The
programming model refers to the paradigm employed by the user in order to
interact with the device. While programmability refers to how the device is
reprogrammed internally. Reprogramming the CSx is hereby decoupled from the
programming performed by the end user.

\subsubsection{Different Degrees of Programmability}

In terms of abstraction, systems are often thought of the be either programmable
or not. However, distinction between individual PFS works would be lacking
using this approach. When more elaborately analyzing PFS different degrees of
programmability emerge. These degrees of programmability can be ordered from
least programmable to most programmable as listed below.

\begin{enumerate}
    \item Transparent operations, (de)compression (Playstation 5 I/O Controller)
    \item Fixed functions, unchangeable, workload specific \cite{2013-fast-active-flash}
    \item Fixed function dataflow programming \cite{Wickremesinghe02distributedcomputing}
    \item Query offloading, SQL, NoSql, Regex \cite{10.14778/2994509.2994512}
    \item Event driven (hooks) user-programmable functions \cite{10.1145/3429357.3430519}
    \item Arbitrary code execution, VHDL, eBPF, TCL \cite{10.1145/605432.605425, kourtis2020safe}
\end{enumerate}

Transparent operations include those such as compression and encryption, this
degree of programmability belongs to fixed function devices.

From here we arrive at fixed functions which are unchangeable and often workload
specific. Typical examples of fixed functions would be database
operations \cite{10.14778/2732967.2732972} or Key-value
operations \cite{10.14778/3137628.3137632, 10.14778/3025111.3025113}. These could
support offloading groupby and aggregate operations or put, get and delete,
respectively.

However, it is not impossible to achieve programmability with fixed functions.
Using a dataflow programming model in combination with small functions called
\textit{functors} general programmability can be achieved. Potentially, this
programmability could even be turing complete.

Even more programmable are the systems using query languages to perform query
offloading. This can be done using both structured and unstructured query
languages such as MySQL and NoSQL.

The next step in programmability is the first to support executing actual user
written code. These event driven systems have programmable hooks. When tied
these hooks run user written code upon activation of an event. Such events
could be read or write I/O operations on the SSD. We typically see this degree
of programmability in dataflow or stream based programming paradigms.

The final degree of programmability is arbitrary code execution, turing complete
programs that can be scheduled and ran on a device at will of the user.

Using these degrees of programmability we can now formally distinquish between
fixed function and programmable flash storage.

\subsection{Evolution of Flash Based Computational Storage}

Before distinguishing between fixed and programmable flash storage, firstly, we
identify all CS works utilizing flash from over the past decade. We present this
overview ordered by publication date. In addition, we show the publication
research field and the state of implementation of the work.

We categorize implementation state into four categories being \textit{proposal},
\textit{simulation}, \textit{prototype} and \textit{deployment}. A work
qualifies for deployment if it is a readily available commercially or has known
real world deployments in commercial systems. Mixes between simulation and
prototype are appropriately attributed based on the distribution between those
two. The overview is shown in table \ref{table:csxoverview}.


\begin{table}[H]
    \caption{Overview of Flash based CSx works ordered by date}
    \centering
    \begin{adjustbox}{width=1\textwidth}
        \begin{threeparttable}[]
            \begin{tabular}{llll}
                \toprule
                \textbf{Name} & \textbf{Publication Date} & \textbf{Research Field} & \textbf{Implementation State} \\
                \midrule
                Active SSD \cite{6062973} & 09-02-2011 & HPC & Prototype \\
                Active Flash \cite{active-flash-piller, 2013-fast-active-flash} & 16-04-2012 & HPC & Simulation \\
                Smart SSD \cite{6558444, 10.1145/2463676.2465295} & 06-05-2013 & HPC & Prototype \\
                Intelligent SSD \cite{10.1145/2464996.2465003, 10.1145/2505515.2507847} & 10-06-2013 & HPC & Simulation \\
                Ibex \cite{10.14778/2732967.2732972} & 15-07-2014 & Databases & Prototype \\
                Willow \cite{186149} & 06-10-2014 & Operating Systems (OS) & Prototype \\
                Biscuit \cite{2016-isca-biscuit} & 18-06-2016 & Computer Architecture (CA) & Prototype \\
                Hadoop ISC* \cite{7524716} & 28-07-2016 & HPC & Prototype \\
                YourSQL \cite{10.14778/2994509.2994512} & 15-08-2016 & Databases & Prototype \\
                Caribou \cite{10.14778/3137628.3137632} & 01-08-2017 & Databases & Prototype \\
                Summarizer \cite{10.1145/3123939.3124553} & 14-10-2017 & Computer Architecture (CA) & Prototype \\
                NDP RE2 regex* \cite{10.1145/3211922.3211926} & 11-06-2018 & Databases & Prototype \\
                Registor \cite{10.1145/3310149} & 26-03-2019 & Storage & Prototype \\
                Cognitive SSD \cite{8839401} & 10-07-2019 & Machine Learning & Prototype \\
                INSIDER \cite{234968} & 15-07-2019 & Storage & Prototype \\
                Catalina \cite{8855540} & 10-08-2019 & HPC & Prototype \\
                THRIFTY \cite{10.1145/3400302.3415723} & 11-02-2020 & Computer Aided Design (CAD) & Simulation \\
                POLARDB \cite{246154} & 14-02-2020 & Storage & Deployment \\
                LeapIO \cite{10.1145/3373376.3378531} & 09-03-2020 & CA / OS & Simulation \\
                CSD 2000 \cite{10.1145/3399666.3399934} & 15-09-2020 & Storage & Prototype \\
                NGD newport \cite{10.1145/3415580} & 12-10-2020 & Storage & Deployment \\
                blockNDP \cite{10.1145/3429357.3430519} & 07-12-2020 & Middleware & Prototype \\
                QEMU CSD* \cite{10.1145/3439839.3459085}  & 26-04-2021 & Storage & Simulation \\ 
                \bottomrule
            \end{tabular}
            \begin{tablenotes}[para,flushleft]
                \centering Overview of selected flash works, their field and
                    implementation state.
            \end{tablenotes}
        \end{threeparttable}
        \label{table:csxoverview}
    \end{adjustbox}
\end{table}

Since early 2020 CS has entered a transnational stage moving from prototypes to
simulations. Here we it seems that we realize that prototypes are not useful if
fundamental problems are not systematically solved using simulations. Excellent
examples are flexible, reconfigurable platforms that allow for fast iterations
such as QEMU CSD \cite{10.1145/3439839.3459085}.

We also see a shift in research fields. Here earlier works had to be disguised 
under HPC while later works are published under storage. This is clear
evidence that CS is now being taken seriously as a part of storage research.

\subsubsection{Fixed Function}

As previously mentioned there is less work being done on fixed function
CS as evidenced by table \ref{table:fixedfunctionoverview}.
Still both of the entries shown are important to the development of CS. The
work on Active Flash \cite{active-flash-piller, 2013-fast-active-flash}
is on of the first to revive CS after research done on Active Disks.

In addition, Caribou \cite{10.14778/3137628.3137632} revives an interesting
concept of providing an alternative interface to the block layer. With
advanced distributed filesystems often using underlying object stores or
key-value stores this might significantly simplify integration of CSxs. This
is further evidenced by the use of an Ethernet based interface as opposed to
SATA or PCIe.

Finally, there is LeapIO that uses a layered offloaded block interface. It uses
local connectivity for offloading or Ethernet with RMDA if the data resides on
another host. The CSx is responsible for accessing this other node over Ethernet
as to not load the host. The host in the evaluation is a guest VM based on
QEMU. More practical applications would be an actual X86 host with PCIe
connected accelerator card.

\begin{table}[H]
    \caption{Fixed function computational storage overview}
    \centering
    \begin{adjustbox}{width=1\textwidth}
        \begin{threeparttable}[]
            \begin{tabular}{lllll}
                \toprule
                \textbf{Name} & \textbf{Programmability} & \textbf{Programming Model} & \textbf{Interface} & \textbf{CSEE} \\
                \midrule
                Active Flash \cite{active-flash-piller, 2013-fast-active-flash} & Fixed functions & N.A. & SATA (OpenSSD) & Embedded \\
                Caribou \cite{10.14778/3137628.3137632} & Fixed functions (key-value store) & Client / Server (RPC) & Ethernet & Bitstream \\
                LeapIO \cite{10.1145/3373376.3378531} & Fixed functions & Transparent & Ethernet (RDMA) & Embedded \\
                CSD 2000 \cite{10.1145/3399666.3399934} & Fixed functions (compression) & Transparent & PCIe (NVMe) & Bitstream \\
                \bottomrule
            \end{tabular}
            \begin{tablenotes}[para,flushleft]
                \centering Overview of identified fixed function CS works across
                    selected literature.
            \end{tablenotes}
        \end{threeparttable}
        \label{table:fixedfunctionoverview}
    \end{adjustbox}
\end{table}

\subsubsection{Programmable Flash Storage}

When looking at PFS research a much larger variety of works is identified.
These are shown in table \ref{table:pfsoverview}. Overall we see early works
using SATA interfaces and Embedded CSEEs while later works mostly use 
PCIe (NVMe) and bitstreams, respectively. Another noteworthy observations is a
steady decline in the dataflow programming model. We suspect the difficulties
from attempting to offload the reduce stage of MapReduce could potentially be
a cause for this decline. In terms of degree of programmability the distribution
is more or less constant with the majority of works being event driven. Second
most common is arbitrary code execution and least is query offloading.

\begin{table}[H]
    \caption{Programmable flash storage overview}
    \centering
    \begin{adjustbox}{width=1\textwidth}
        \begin{threeparttable}[]
            \begin{tabular}{lllll}
                \toprule
                \textbf{Name} & \textbf{Programmability} & \textbf{Programming Model} & \textbf{Interface} & \textbf{CSEE} \\
                \midrule
                Active SSD \cite{6062973} & Event driven & Dataflow (streams) & PCIe & Operating system (Custom) \\
                Smart SSD \cite{6558444} & Event driven & Dataflow (MapReduce) & SATA & Embedded \\
                Smart SSD \cite{10.1145/2463676.2465295} & Event driven & Shared memory & SATA & Embedded \\
                Intelligent SSD \cite{10.1145/2464996.2465003, 10.1145/2505515.2507847} & Arbitrary code execution\footnotemark[7] & Shared memory\footnotemark[7] & N.A. & Operating system (Linux)\footnotemark[7] \\
                Ibex \cite{10.14778/2732967.2732972} & Query offloading (MySQL) & Declarative & SATA & Bitstream \\
                Willow \cite{186149} & Arbitrary code execution & Client / Server (RPC) & PCIe (NVMe) & Operating system (Custom) \\
                Biscuit \cite{2016-isca-biscuit} & Event driven & Dataflow & PCIe & Embedded \\
                Hadoop ISC* \cite{7524716} & Event driven & Dataflow (MapReduce) & SAS & Embedded \\
                YourSQL \cite{10.14778/2994509.2994512} & Query offloading (MySQL) & Declarative & PCIe (NVMe) & Bitsream\footnotemark[8] \\
                Summarizer \cite{10.1145/3123939.3124553} & Event driven & Shared memory & PCIe (NVMe) & Embedded \\
                NDP RE2 regex* \cite{10.1145/3211922.3211926} & Query offloading (Regex) & N.A. & N.A. & Embedded \\
                Registor \cite{10.1145/3310149} & Query offloading (Regex) & Shared memory & PCIe (NVMe) & Bitsream \\
                Cognitive SSD \cite{8839401} & Arbitrary code execution & Shared memory & PCIe (NVMe, OpenSSD) & Accelerators (Custom) \\
                INSIDER \cite{234968} & Event driven & Shared memory (VFS) & PCIe & Bitstream \\
                Catalina \cite{8855540} & Arbitrary code execution & Client / Server (MPI) & PCIe (NVMe) & Operating system (Linux) \\
                THRIFTY \cite{10.1145/3400302.3415723} & Event driven\footnotemark[9] & Shared memory (VFS)\footnotemark[9] & PCIe\footnotemark[9] & Bitstream\footnotemark[9] \\
                POLARDB \cite{246154} & Query offloading (POLARDB) & Declarative & PCIe & Bitstream \\
                NGD newport \cite{10.1145/3415580} & Arbitrary code execution & Client / Server & PCIe (NVMe) & Operating system (Linux) \\
                blockNDP \cite{10.1145/3429357.3430519} & Event driven & Dataflow (streams) & PCIe (NVMe, OpenSSD) & Virtual Machine (QEMU) \\
                QEMU CSD* \cite{10.1145/3439839.3459085} & Arbitrary code execution & Shared memory & PCIe (NVMe) & N.A. (Simulated) \\
                \bottomrule
            \end{tabular}
            \begin{tablenotes}[para,flushleft]
                \centering Overview of PFS works and various aspects as
                    previously detailed.
            \end{tablenotes}
        \end{threeparttable}
        \label{table:pfsoverview}
    \end{adjustbox}
\end{table}

\footnotetext[7]{Simulations performed by porting workloads unto ARM based
processor. No actual hardware on SSDs is used.}

\footnotetext[8]{The work uses special FCPs with hardware based filtering
functions. We assume these must be implemented using FPGAs although the work
does not specify.}

\footnotetext[9]{Build on top of the INSIDER software stack.}

Other more specific observations on individual works include those on the use of
the MapReduce framework. None of the implementations fully offload each stage.
Typically only the mapping stage is offloaded as it has no cross data
dependencies. This is unfortunate as the reduce stage would offer the highest
reduction in data transferred. Interestingly this is contrarily to the claims of
Hadoop ISC \cite{7524716} that argues most data reduction happens in the
\textit{mapping} stage.

In terms of query offloading the work on POLARDB \cite{246154} is interesting as
it sees actual deployment in Alibabas private cloud infrastructure. This is
the only PFS deployment we see across all selected works together with
NGD newport \cite{10.1145/3415580}. The newport system is a Flash based CSD that
supports a wide range of programming models by offering SSH access to the
device alongside some client / server based APIs. However, although very
flexible we do not think this architecture will lead to widespread adoption as
we will explain in a follow up section.

Other interesting architectures include that of Catalina \cite{8855540} which
allows programming using the Messaging Passing Interface (MPI). Similarly, we
do not expect widespread adoptions as this approach highly complicates data
placement and access.

Out of all these works only tried to overcome the traditional block interface by
utilizing OpenSSD. This new storage interface directly exposes the underlying
flash architecture significantly reducing the semantic gap between host and
device. When done the Flash Translation Layer (FTL) becomes an integrated part
of the host as opposed to the device. We call these type of interfaces
host-managed. A successor to OpenSSD will be addressed in a following section.

While all these works fail to address some fundamental issues which hinder
adoption at least the approach of QEMU CSD \cite{10.1145/3439839.3459085} allows
for rapid prototyping using its simulation framework. Unfortunately, the source
code of this simulation platform is not made available. This makes
reproduction of their results impractical as well making this rapid prototyping
platform inaccessible to others.

Clearly we have seen advancements in flash based CS over the last decade. From
early basic prototypes with fixed workload specific functions tied to the
FTL \cite{active-flash-piller, 2013-fast-active-flash}. To know commercial use
of transparent decompression in the Playstation 5 its I/O controller.

In PFS we also see progression as many prototypes now have led to at least two
cases of commercial use. We also see the initial developments for an accelerator
like CSEE similar to Vulkan or OpenCL in CognitiveSSD \cite{8839401}. We believe
this type of architecture has a high change of achieving widespread adoption. In
the next section we will address key challenges that remain for future
developments.

\subsection{Complexity and Challenges}

Several key-challenges remain throughout the past decade of flash based CS. We
see two fundamental works identifying these
\cite{10.1145/3102980.3102990, barbalacecomputational}. In addition we provide
some of our own. Unfortunately, it is not possible to go in to depth identifying
how each work of the previous section fairs regarding each challenge. This
in depth analysis is left as future work.

Firstly, the work by Barbalace et al \cite{10.1145/3102980.3102990} identify
some of the challenges that require explicit support. These are
\textit{locality}, \textit{protection}, \textit{scheduling},
\textit{programmability} and \textit{low-latency}. The work does not address
if this support is to be implemented by the host operating system or the device
apart from scheduling. Here it argues that scheduling should be implemented by
the device and we agree. The two primary challenges that remain largely unsolved
are \textit{protection} and \textit{low-latency}. However, addressing either of
these challenges might additionally complicate \textit{scheduling}.

The subsequent work by the same authors describes \textit{resource
management}, \textit{security}, \textit{data consistency} and
\textit{usability} as open research questions \cite{barbalacecomputational}.
Some of these have clear overlap such as \textit{security} and
\textit{protection}.

However, this overlap is logical as one of these challenges are from the
perspective of what requires additional changes to existing operating
systems\footnotemark[10]. While they other aims to identify open research
questions which hinder adoption.

\footnotetext[10]{With operating system being used to refer to both the overall
software on the host and the device in the context of this work.}

Across all these open research questions hindering adoption we see one clear
theme. The questions resolve around trying to share information between the
host and device. Be it replication maps or filesystem information. We also see
this in the different programming models as described in the same work.
Similarly we see this problem appear when trying to offload the reduce stage of
MapReduce to CSxs \cite{7524716}.

This sharing requirement is natural given that CS requires complete vertical
integration. From nand flash interface, FPGA / CPU design, programming
methodology, peripheral interface, hardware abstraction layer and user
programming interface. Across this entire stack changes are needed.




However, we argue that combining block filesystem access with CS introduces
unfeasible levels of complexity in the FTL of the device. Just like two
filesystem partitions wont share overlapping storage space, partitions and CS
should not share the same storage space. Research should instead focus on using
CS platforms to create virtual filesystems on top of them. This is similar to
CephFS that moved away from using XFS as underlying filesystem. Instead CephFS
now uses BlueStore object storage which uses the whole storage device without a
filesystem. Additionally, we already see this approach in a recently proposed
CS key-value store \cite{273719}. In addition, moving away from this requirement
already lead to some works achieving real world deployments \cite{246154}.

We propose to abolish the idea of sharing fundamental information between the
device and host, the attempt to bridge The semantic gap. Instead all effort
should be focused on developing CSx interfaces and APIs and how to develop host
features such as virtual file systems on top of those.

But most important, even more so than the open research questions, is the lack
of open-research causing many works to have to reinvent the wheel. The lack of
open-source hardware and software design in the field not only severely hinders
reproducibility but also the speed at which the field advances. Lack of access
to the designs or software of previous works requires new research to often
start from scratch. In our opinion any work utilizing newly written software or
hardware designs should release these under an open-source license or be
rejected for publication period.

\section{Future Prediction}


Full blown heterogeneous architectures will become the
norm \cite{10.1145/3102980.3102992, 10.1145/3458336.3465291}. We already see this
transition in devices such as iPhones. The use of neural accelerators is
prevalent in modern mobile SoCs. More recently, we see a push for neural
accelerators in datacentres as well. It is only a matter of time before we start
seeing these appear in consumer desktop platforms as well.

\subsection{Interface Evolution}

To support such a heterogeneous architectures several interfaces must evolve.
One of these interfaces is the block device interface which will gradually
dissapear from flash based devices \cite{death-interface}. We will see two
prominent types of interfaces emerge for flash based devices. The first being
a key-value store with offloaded scan and gather functionality. The second being
an accelerator interface much like Vulkan and OpenCL.

The overall adoption of these accelerator interfaces depends on our performance
as software and hardware engineers. We must design a good performing system with
easy integration into existing operating systems and programs. The success of
this interface must not be tied to any specific programming model supporting
dataflow, shared memory and client / server as minimum. The hardware abstraction
layer will use a dataflow like programming model. While the support for shared
memory is built on top of this dataflow model. Lastly, the client / server model
is in turn built on top of the shared memory model.

Using these variety of programming models we will see operating systems built
high level abstractions such as virtual filesystems. In addition we will see
specialized applications such databases stored entirely on one or multiple CSxs.

Naturally, presenting an entire SSD for a single purpose can be undesirable.
To alleviate this new NVme interface extensions like Zoned Namespaces (ZNS)
will be relied upon heavily. While ZNS still presents a block like interface it
much better represents the underlying nand Flash architecture eliminating
complexity in the FTL, excessive garbage collection and write amplification.

ZNS will be the predominant interface for CS until a complete transition away
from block like interfaces is made. There is already significant literature
suggesting the adoption of ZNS \cite{liu_2021, 10.1145/3448016.3457540,
10.1145/3458336.3465300, 273945}. With already the first CS work utilizing it
as well \cite{273709}.

Beyond storage interfaces such as ZNS we will see more and more use of
interconnects being used across various components in desktop and server
systems. Some of these will be cache coherent such as
OpenCAPI \cite{opencapi_interconnect} or CCIX \cite{ccix_interconnect}. Others
aimed specifically at communication for accelerators such as
GenZ \cite{genz_interconnect} and CXL2 \cite{cxl2_interconnect}.

No CS will not depend on if it makes sense from an energy efficient or
architectural viewpoint to use. Instead it relies entirely on the ease of
integration, reliability, predictability and performance of the overall systems
we manage to design.

\subsection{The Arms Race}

Both the throughput of nand flash as well as the bandwidth of interconnects such
as PCIe will increase over the coming years. There will be no clear winner
which sets the president for requiring a well designed CS system if we want to
see adoption. PCIe will start using realtime compression, similar to how HDMI
2.1 support Display Stream Compression (DSC). The maximum permissible length
of PCIe traces as well as DisplayPort and HDMI cables will decrease further and
further. Meanwhile the cost of manufacturer of components such as motherboards
will increase. This is due to the signal integrity requirements posed by such
high-speed differential signaling.

Eventually this arms race will reach a tipping point until the use of onboard
fiber interconnects becomes the norm. These fiber channels will be integrated
into regular PCBs occupying only several layers of the PCB stack. This will
allow for seemingly unlimited bandwidth with nearly no signal degradation and
no cross talk. In the near future, copper based differential signaling will die.

\section{Future Work}

One key element of a literature survey is the methodology used to identify,
categorize and analyze related literature. Future work could better address
certain characteristics of our approach. Key areas for improvement being
accuracy and fairness.

Firstly, plainly assigning keywords to works based on the search term used to
find them is naïve. A better approach would categorize works based on the
keywords appearing in title and abstract.

Secondly, the selection procedure is critical when dealing with a limited number
of selected literature. Analyzing works ordered by publishing date introduces
unwanted bias. This bias will favor selecting more historical works over recent
contributions. This is due to a steadily approaching hard limit on the amount of
works that can still be selected. A better approach would analyze works from the
initial selection at random.

Lastly, the assessment of novelty can be substantially improved to account for
the different phases of a research field. This is because preliminary research
in a field has a tendency to appear more novel than later works. As example, the
first implementation of a PFS architecture for database query offloading can
appear more novel than subsequent works. These subsequent works could address
multi-user tenancy both being equally important to progress the field. A better
approach should separate the works in different phases based on date. As well as
identifying the current open research question in later phases of the field.
The assessment in later phases should than evaluate if any of the open research
questions are being addressed.

Three works stand out that would potentially be included should our approach
have been improved as proposed. These are NASCENT \cite{10.1145/3431920.3439298},
FERMAT \cite{9444075} and COPRAO \cite{9438805}.

Beyond improvements to our approach we see several potential topics that might
suit other literature surveys. First is relating the development of CS to
support technologies such as ZNS or OpenCAPI. While we addressed such support
technologies briefly we feel much more information can be derived here.
Similarly, the advancement of open standards and ISAs should form an interesting
topic with many recent contributions.

In addition there are several areas of our work that could be easily expanded.
One area is the ubiquity of computational storage. Due to constrained amounts of
literature that could be selected this part of our work is not nearly as
exhaustive as it could be. In addition, we have identified many different fixed
and programmable flash based CSxs. However, there is much more information that
could be compared across these works. These comparisons include the integration
of filesystem support or lack there of. Moreover, how transparent the offloading
is to the end user or how much energy and performance improvement the work
constituted.

Clearly there is plenty of research opportunity and exciting new technologies
on the horizon for the field of CS. Will you help us make it a success?

\section{Glossary}

\begin{enumerate}
    \item CPU - Central Processing Unit
    \item DRAM - Dynamic Random Access Memory
    \item PCIe - Peripheral Component Interconnect Express
    \item CS - Computational Storage
    \item CSx - Computational Storage Device
    \item PFS - Programmable Flash Storage
    \item PFSD - Programmable Flash Storage Device
    \item OCSSD - Open Channel SSD
    \item ZNS - Zoned Namespaces
    \item HAL - Hardware Abstraction Layer
	\item ACL - Access Level Control
    \item CSF - Computational Storage Function
    \item FPGA - Field Programmable Gate Array
    \item DSP - Digital Signal Processor
    \item DAG - Directed Acyclic Graph
    \item NIC - Network Interface Controller
    \item PIM - Programmable in Memory
    \item CSP - Computational Storage Processor
    \item CSA - Computational Storage Array
    \item CSD - Computational Storage Driver
    \item CSE - Computational Storage Engine
    \item CSEE - Computational Storage Execution Environment
    \item CSR - Computational Storage Resource
    \item CSF - Computation Storage Function
    \item FDM - Function Data Memory
    \item ADFM - Allocated Function Data memory
    \item HAL - Hardware Abstraction Layer
    \item ICD - Installable Client Driver
    \item FTL - Flash Translation Layer
    \item MPI - Message Passing Interface
    \item VLIW - Very Long Instruction Word
    \item LUN - Logical Unit
	\item Upstream - When a certain patch or feature has been merged into the
					 main repository of a project.
	\item GUI - Graphical User Interface
	\item IDE - Integrated Development Environment
	\item Target - A target is something that can be compiled or generated and
				   sometimes executed. Examples include PDFs using LaTeX,
				   binaries or libraries.
	\item Binary - A binary is a file containing machine instructions that can
				   be directly executed. A compiled C file containing a main
				   method is a binary while a shell script is not. Typically
				   these binaries contain metadata information to aid the
				   underlying operating system in executing them, a common
				   format for this metadata is ELF.
	\item ABI - Application Binary Interface.
	\item API - Application Programming Interface.
	\item SCSI - Small Computer System Interface
	\item ATA - AT Attachment
	\item ZBC - (SCSI) Zoned Block Command
	\item ZAC - (ATA) Zoned ATA Commands
\end{enumerate}

\bibliographystyle{ACM-Reference-Format}
\bibliography{bibliography}

\end{document}